\documentclass[prl,aps,twocolumn,showpacs,superscriptaddress,amsmath,amssymb,floatfix,reprint]{revtex4}
\usepackage{graphicx}
\usepackage{dcolumn}
\usepackage{bm}
\usepackage{float}
\usepackage{hyperref}

\newcommand{\nodag}{{\phantom{\dag}}}
\newcommand{\kk}{{\bf k}}
\newcommand{\qq}{{\bf q}}

\newcommand{\kkA}{{\kk_A}}
\newcommand{\kkN}{{\kk_N}}

\newcommand{\Ek}{E_\kk}
\newcommand{\uk}{u_\kk}
\newcommand{\vk}{v_\kk}
\newcommand{\vvk}{{\bf v_k}}

\newcommand{\eps}{\varepsilon}
\newcommand{\epsk}{\varepsilon_\kk}

\newcommand{\Deltak}{\Delta_\kk}

\newcommand{\gammadag}{\gamma^\dag}
\newcommand{\gammaks}{\gamma_{\kk\sigma}^{\nodag}}
\newcommand{\gammaksdag}{\gamma_{\kk\sigma}^{\dag}}

\begin{document}

\title{Resonant Inelastic X-ray Scattering as a Probe of the Phase and Excitations of the Order Parameter of Superconductors}
\author{Pasquale Marra}
\affiliation{Institute for Theoretical Solid State Physics, IFW Dresden, D-01069 Dresden, Germany}
\author{Steffen Sykora}
\affiliation{Institute for Theoretical Solid State Physics, IFW Dresden, D-01069 Dresden, Germany}
\author{Krzysztof Wohlfeld}
\affiliation{Institute for Theoretical Solid State Physics, IFW Dresden, D-01069 Dresden, Germany}
\author{Jeroen van den Brink}
\affiliation{Institute for Theoretical Solid State Physics, IFW Dresden, D-01069 Dresden, Germany}
\affiliation{Department of Physics, TU Dresden, D-01062 Dresden, Germany}



\begin{abstract}

The capability to probe the dispersion of elementary spin, charge, orbital, and lattice excitations has positioned resonant inelastic s-ray scattering (RIXS) at the forefront of  photon science. Here we develop the scattering theory for RIXS on superconductors, calculating its momentum-dependent scattering amplitude. Considering superconductors with different pairing symmetries we show that the low-energy scattering is strongly affected by the superconducting gap and coherence factors. This establishes RIXS as a tool to disentangle pairing symmetries and to probe the elementary excitations of unconventional superconductors.

\end{abstract}

\maketitle

\paragraph{Introduction}

In the past decade, resonant inelastic x-ray scattering (RIXS)~\cite{Schuelke2007,Ament2011} 
has made remarkable progress as a spectroscopic technique, establishing itself 
as an experimental probe of elementary 
spin~\cite{Braicovich2009a,Braicovich2009b,Hill2008,Tacon2011,Kim2012b,Dean2012}, 
orbital~\cite{Ulrich2009,Schlappa2012}, 
and lattice excitations~\cite{Yavas2010}. 
In quite a number of cases, theoretical considerations  have preceded and stimulated these experimental 
advances, prominent examples being the theoretical demonstration of the presence of 
strong single-magnon scattering channels in cuprates~\cite{Groot1998,Ament2009a} and 
iridates~\cite{Ament2010b}. 
Being a photon-in/photon-out spectroscopy, both the energy $\hbar \omega$ and the momentum change 
${\bf q}$ of the scattered photon are measured. As the energy and momentum lost by the 
photon are transferred to intrinsic excitations of the material under study, direct 
information on the dispersion of those excitations becomes available. The resonant character of the 
technique is due to the energy of the incident photon being chosen such that it coincides, and hence resonates, with an x-ray absorption edge of the system~\cite{Schuelke2007,Ament2011}. 
This year the energy resolution of RIXS has reached $\sim$30 meV in the hard x-ray 
regime~\cite{Kim2012b}, 
will reach 50 meV in the soft x-ray regime by building on present instrumentation~\cite{PrivComm} and is designed to reach 11 meV at the Cu $L$-edges at the NSLS-II presently under construction~\cite{NSLS-II_design_report}. 
This brings the RIXS energy resolution well into the regime of the energy gap of cuprate superconductors, which 
stretches out to 119 meV for mercury-based high $T_c$ systems~\cite{Yu2009}.
Consequently the fundamental question arises of how the superconducting (SC) state 
leaves its fingerprints in RIXS spectra --- in particular whether and how RIXS is sensitive to 
the phase and amplitude of the SC gap and to quasiparticle excitations. 

Probing the order parameter in unconventional superconductors is generally the first 
step for an investigation of the pairing mechanism and of the character of the SC state. 
Compared to the available 
spectroscopic methods, such as scanning-tunneling spectroscopy (STS), photo-emission spectroscopy, optical 
spectroscopy or inelastic 
neutron scattering, RIXS uniquely combines the advantages of bulk-sensitivity and 
availability of momentum resolution while at the same time requiring only small sample volumes. 
Here we show how the sensitivity of the RIXS process to the dynamical structure 
factor (DSF) of the electron spin and density in the SC state enables the 
investigation of SC quasiparticle 
excitations. In particular, we find that the momentum dependence of RIXS spectra 
is intrinsically determined by the pairing symmetry, being sensitive not only to 
the magnitude of the SC gap and to the presence of 
nodes on the Fermi surface but also to the phase of the order parameter. 
This phase sensitivity is due to the appearance of SC coherence factors which, for 
instance, in STS determine to 
large extent the quasiparticle interference in the presence of 
impurities~\cite{Hoffman2002,McElroy2003,Hanaguri2007,Kohsaka2008,Hanaguri2009,Hanke12,Sykora11}.

\paragraph{Dynamical structure factors in RIXS}
In what follows we concentrate on the so-called \emph{direct} RIXS process 
at the transition metal (TM) ion $L$ edges, in which the incoming photon 
resonantly excites the core shell 2$p$ electron into the 3$d$ shell which 
consequently decays into an outgoing photon and a charge, spin,
or orbital excitation in the electronic system~\cite{Ament2011}. 
In this case the RIXS cross section can be written as~\cite{Ament2009a, Ament2011, Haverkort2010, Marra2012}
\begin{align}
\label{eq:crosssection}
I_{\bf e}({\bf q},\omega)= \sum_f \big| \langle f | \hat{O}_{{\bf q},{\bf e}} 
| i \rangle \big|^2 \delta (\hbar \omega - E_f + E_i),
\end{align}
where $|i\rangle$ ($|f\rangle$) is the initial (final) state of RIXS process with energy $E_i$ ($E_f$) 
and $\hbar \omega$ (${\bf q}$) is the transferred photon energy (momentum).
Furthermore, in the fast collision approximation (FCA)~\cite{Ament2009a, Ament2011, Haverkort2010, Marra2012}
the Fourier transformed RIXS transition operator $\hat{O}_{{\bf q},{\bf e}}$ can be
written as $\hat{O}_{{\bf q},{\bf e}}=W^c_{\bf e} \rho_{\bf q} +  {\bf W}^s_{\bf e}\cdot{\bf S}_{\bf q}$, 
where the operators $\rho_{\bf q}=\sum_{\kk\sigma} d^{\dagger}_{\kk+\qq\sigma} d^{}_{\kk\sigma}$ and ${\bf S}_{\bf q}$ 
are the density and spin of conduction electrons~\cite{Haverkort2010, Marra2012, Note1}.
The so-called RIXS form factors $W^{c}_{\bf e}$ and ${\bf W}^{s}_{\bf e}$ depend 
on the TM ion, 
the specific geometry of the experiment, and on the polarization ${\bf e}$ of the incoming 
and outgoing photon --- their
precise dependencies are provided in Refs.~\onlinecite{Ament2009a, Haverkort2010, Marra2012}.
In terms of these form factors the RIXS cross section is
\begin{align}\label{eq:crosssectionb}
I_{\bf e}({\bf q}, \omega) = |W^c_{\bf e}|^2 \chi^c ({\bf q}, \omega) 
+ |{\bf W}^s_{{\bf e}}|^2 \chi^s ({\bf q}, \omega),
\end{align}
where $\chi^c=\sum_f | \langle f | {\rho}_{{\bf q}} | i \rangle|^2 
\delta (\hbar \omega - E_f + E_i)$ 
is the charge and $\chi^s = \sum_f | \langle 
f | S^z_{{\bf q}} | i \rangle|^2 \delta (\hbar \omega - E_f + E_i) $
the spin dynamical structure factor (DSF), where one assumes that the spin DSF has the same momentum and energy dependence for all the 
three different components of the spin operator --- as is the case of 
unconventional superconductors~\cite{Andersen2005}.
As is clear from the above, the amplitude of the RIXS form factors $W^{c}_{\bf e}$ and ${\bf W}^{s}_{{\bf e}}$
can be tuned by properly adjusting the experimental conditions in RIXS. This implies that RIXS at $L$-edges 
can measure either spin or charge DSF depending on the chosen polarization, which is a unique feature of RIXS spectroscopy.
Note that the FCA has been successfully used to describe low energy excitations in RIXS, e.g.,
at the Cu $L$ edge of both undoped and doped cuprates
(cf. theoretical calculations~\cite{Ament2009a, Haverkort2010, Marra2012} and experimental results with their interpretation fully based on the FCA~\cite{Braicovich2010, Tacon2011, Dean2012, Schlappa2012}),
at the Fe $L$ edge of the iron arsenides~\cite{Kaneshita2011},
and at the Ir $L$ edge in iridates~\cite{Kim2012a}. 
It has also been shown theoretically that when the incoming photon energy in RIXS is tuned to a TM
ion resonant edge in a TM oxide, the FCA describes the RIXS spectrum well~\cite{Haverkort2010}.

In the following we concentrate on determining the properties of DSF for 
different types of singlet-pairing superconductors, 
and to be even more specific below we consider the case of Cu ions in lattices with tetragonal symmetry, 
i.e., the one which has direct relevance to the high $T_c$ superconductors. 
Our main aim in this context is to establish how
a variation of the phase of the SC order parameter is reflected in the
spin and charge DSF.
Following the most direct theoretical inroad and avoiding model-specific technical details, we
performed calculations for a 
singlet-pairing superconductor with a SC order parameter varying 
along the Fermi surface. 
Even if by this electron correlations are not fully taken into account, this approach is commonly used --- and is very successful
to calculate quasiparticle interference in cuprates~\cite{Hanaguri2007,Hanaguri2009,Kohsaka2008,Fischer2007}.
Besides this, in Appendix~III\ref{app:strong} we show that it is actually possible to introduce strong correlations between electrons
into the calculations and that such does not affect the main results on the SC electronic system presented below. 

\paragraph{DSF for a superconductor}

Quasiparticle excitations in a single band 
superconductor with singlet-pairing 
are described by the Hamiltonian
$H-\eps_F N=\sum_{\bf k} \Ek~\gammaksdag \gammaks$,
where $E_\kk=\sqrt{\epsk^2+\vert\Deltak\vert^2}$ is the quasiparticle 
energy dispersion depending on the SC gap function $\Deltak$ and on the 
dispersion of the bare electrons $\epsk$. The Bogoliubov quasiparticle 
operators $\gammaks$ are related to the 
$d$ electron operators via $d_{\kk\uparrow}=\uk^* \gamma_{\kk\uparrow}-\vk 
\gammadag_{-\kk\downarrow}$ and $d_{\kk\downarrow}=\uk^* 
\gamma_{\kk\downarrow}+\vk \gammadag_{-\kk\uparrow},$
with $\vert\uk\vert^2(\vert\vk\vert^2)=\frac12\left(1\pm\epsk/\Ek\right)$  
and $\uk^*\vk=\frac12\Deltak/\Ek$. The Bogoliubov transformation allows 
the evaluation of the DSF for a SC by evaluating the
transition amplitudes $\langle f \vert \rho_{\bf q}\vert i \rangle$ 
and $\langle f \vert S^z_{\bf q}\vert i \rangle$ between 
the ground state $|i \rangle$ and any excited state $|f\rangle$ 
of the Hamiltonian. At zero temperature, the excited states which 
contribute to DSF have the form  
$\gammadag_{\kk+\qq,\sigma}\gammadag_{-\kk,-\sigma}|g\rangle$ 
with a transition energy of $E_{\kk+\qq}+E_{-\kk}$. Using the 
Bogoliubov transformation one then finds that the DSF for a 
superconductor reads
\begin{align}\label{StructureFactor}
\chi^{c, s} ({\bf q}, \omega) = \sum_\kk & \left[ 1 \pm \frac{{\rm Re} 
(\Deltak^{}\Delta_{\kk+\qq}^*)\mp\epsk\eps_{\kk+\qq}}{E_\kk E_{\kk+\qq}} \right] \nonumber \\
& \times \delta(\hbar\omega-E_{\kk+\qq}- E_\kk)
\end{align}
where $\pm$ sign is for the charge (spin) DSF~\cite{Kee1998, Kee1999, Voo2000}.
Thus, the DSF  is a sum over all 
momenta within the Brillouin zone,
where the transition amplitudes are strongly 
influenced by the 
character of the SC state. 
Although quasiparticle interactions substantially affect the DSF, 
they do not alter its intrinsic sensitivity to the character 
of the SC state (in particular to the symmetry of its gap function), 
as can be seen, e.g., at the random-phase approximation level~\cite{Kao2005}, 
or by considering a strongly correlated system with Hubbard interactions, see Appendix~III\ref{app:strong}.
\begin{figure}
\centering
\includegraphics[width=1\columnwidth]{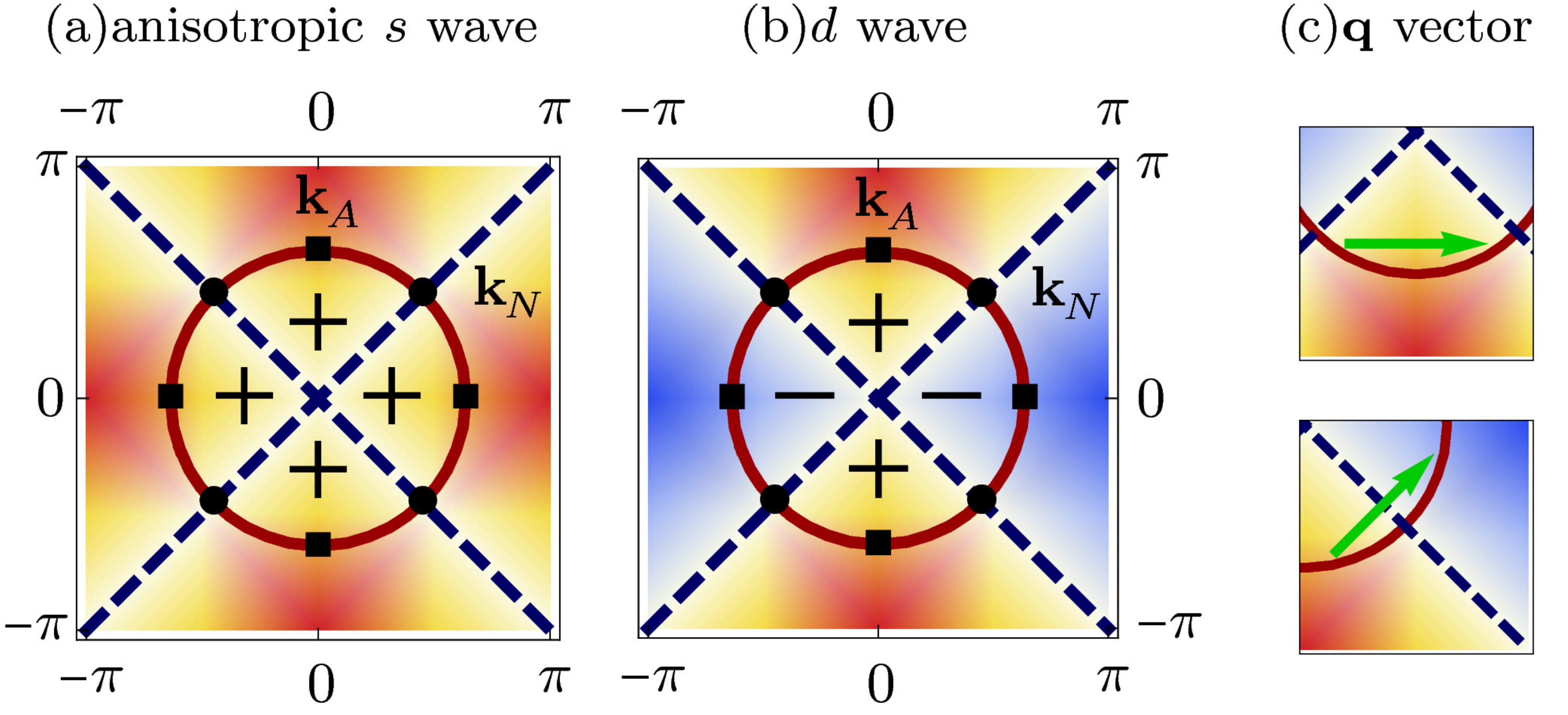}
\caption
{
(color online) Order parameters of an anisotropic 
$s$ wave (a) and $d$ wave (b) superconductor with an isotropic Fermi surface 
(solid line). 
The order parameter vanishes (has maxima) at the nodal points $\kkN$ (anti-nodal points $\kkA$). 
(c) Particle-hole excitations with and without sign reversal in case of $d$ wave pairing.
}
\label{fig:OrderParameter}
\end{figure}

To determine in detail how the RIXS spectra of 
unconventional superconductors reflect the phase of the order parameter, 
we consider (i) a $d$-wave pairing with 
$\Delta_{\bf k} \propto (\cos k_x - \cos k_y)$ and (ii) an anisotropic 
$s$-wave pairing with $\Delta_{\bf k} \propto |\cos k_x - \cos k_y|$, i.e.,
two pairing symmetries which differ from each other only in the SC order parameter phase. 
Maps of the considered gap functions and the Fermi surface are shown 
in Fig.~\ref{fig:OrderParameter}. In the $s$ wave case, the DSF 
$\chi^{c} ({\bf q}, \omega)$, due to the '$+$' sign in 
Eq.~\eqref{StructureFactor}, is non-zero all over the Brillouin zone, 
while in the $d$ wave case excitations combining 
momenta with opposite phases of the order parameter, i.e.,
sign reversing processes, are suppressed. Note that for the spin 
DSF this situation is inverted.

\begin{figure}
\centering
\includegraphics[width=1\columnwidth]{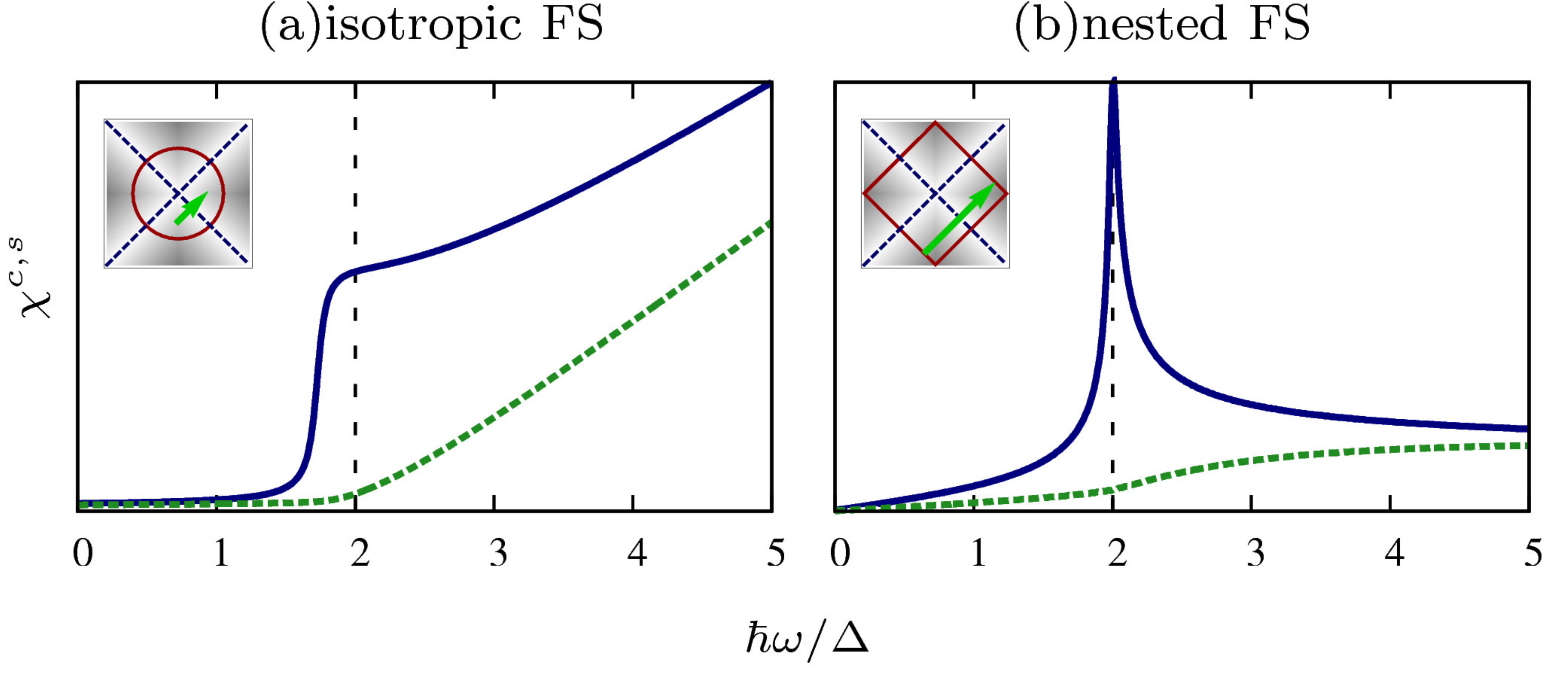}
\caption
{
(color online) Dynamical structure factor (DSF) in a SC for an isotropic (a) 
and for a nested Fermi surface (b) with transferred momentum 
$\qq=(k_F,k_F)/\sqrt{2}$ and $\qq=(\pi,\pi)$, for 
charge excitations ($\chi^c$) with anisotropic $s$ wave (solid line) and $d$ wave (dashed line), 
and 
for spin excitations ($\chi^s$) with $d$ wave (solid line) and 
anisotropic $s$ wave (dashed line) order parameter.
}
\label{fig:PhaseCoherence}
\end{figure}

\paragraph{Phase sensitivity}
The calculated quasiparticle spectra for the above two pairing symmetries
are shown in Fig.~\ref{fig:PhaseCoherence}(a) for a fixed and relatively large
momentum transfer $q=k_F$, where $k_F$ is the Fermi radius. 
A detailed expansion of the DSF for $q \ll k_F$ is provided in Appendix~I\ref{app:smallq}.
Note that direct RIXS at the $L$-edge in 2D cuprates allows 
momentum transfers $q\lesssim 0.87 \pi$~\cite{Braicovich2010}
and therefore one is able to access momentum 
transfers $\bf q$ of the order of $k_F$. Such momentum transfers correspond 
to momentum vectors combining two anti-nodal points ${\bf k}_A$ on the 
Fermi surface [see inset of Fig.~\ref{fig:PhaseCoherence}(a)] 
with the same gap value $\Delta_{{\bf k}_A} = \Delta$. 
For such an excitation, the sign of the order parameter in the $s$ wave case is preserved
whereas in the $d$ wave case it is reversed. 
Clearly the spectral weight at $\hbar \omega=2|\Delta|$ 
in the charge DSF is enhanced in the $s$ wave case. On the other hand, for the $d$ wave 
case the sign reversal leads immediately to a suppression of the 
DSF according to Eq.~\eqref{StructureFactor}. Therefore, the spectral 
weight at $\hbar \omega=2|\Delta|$ is expected to be very small which 
is confirmed by the dotted curve in Fig.~\ref{fig:PhaseCoherence}(a). 
On the other hand, spin DSF is suppressed in the $s$ wave case, 
while the sign reversal enhances the spectral weight at $\hbar \omega=2|\Delta|$.
Note that due to the gap being equal in magnitude for both cases, the 
obtained effect is entirely due to phase changes of the SC order parameter 
along the Fermi-surface.

In unconventional superconductors, where the pairing is generally considered as mediated 
by antiferromagnetic spin fluctuations~\cite{Schrieffer1989}, the SC gap 
function is expected to exhibit a sign reversal between 
the Fermi momenta connected by characteristic wave vector ${\bf Q}$ of the spin 
fluctuations~\cite{Kuroki2001}. The conduction electrons of such 
superconductors show a tendency to Fermi-surface nesting  with a typical 
nesting vector of ${\bf Q} = (\pi,\pi)$.  In Fig.~\ref{fig:PhaseCoherence}(b)  
the scattering intensity as function of energy for the two
pairing symmetries considered above are shown again, but now for a perfectly 
nested cuprate-like Fermi surface (see inset),
with a transferred momentum equal to the nesting vector.
It is clear that, in comparison to the case of an isotropic Fermi surface, the 
coherence peak in the anisotropic $s$ wave pairing case appears now 
strongly enhanced due to the nesting effect. 
As in the case of an isotropic Fermi surface, 
sign reversing excitations occurring in the $d$ wave case in the charge DSF
are strongly suppressed, as well as
sign preserving excitations in the anisotropic $s$ wave case in the spin DSF.

\begin{figure}
\centering
\includegraphics[width=1\columnwidth]{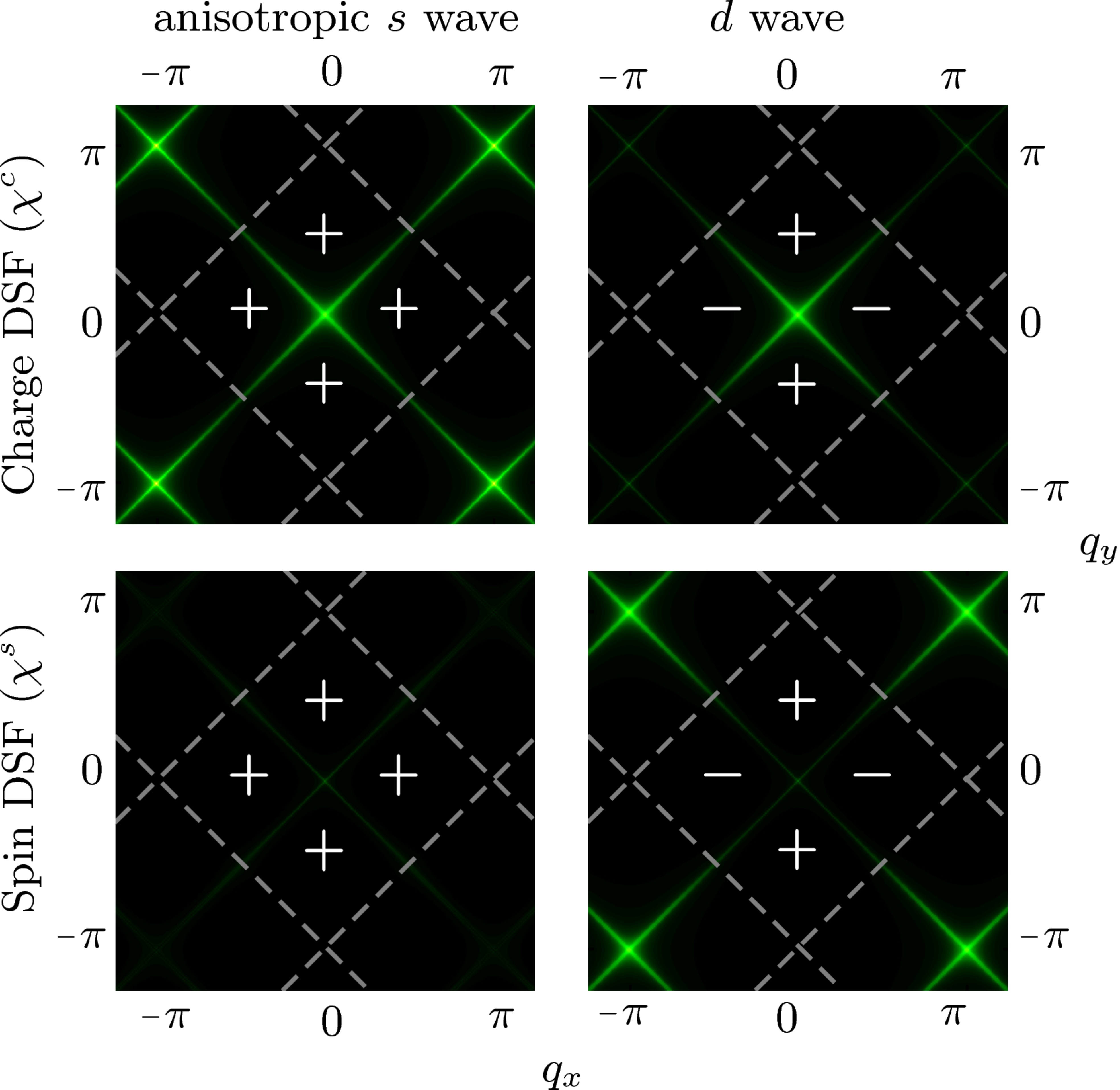}
\caption
{
(color online) 
Charge ($\chi^c$) and spin ($\chi^s$) DSF at fixed energy 
$\hbar \omega=2|\Delta|$, as a function of transferred momentum 
for an anisotropic $s$ wave and a $d$ wave SC with a nested Fermi surface.
}
\label{fig:PhaseBZ}
\end{figure}

To highlight its strong dependence on the order parameter phase, 
we show in Fig.~\ref{fig:PhaseBZ} the DSF for a fixed energy $\hbar \omega=2|\Delta|$ 
as a function of momentum ${\bf q}$
in the entire Brillouin zone, both for the anisotropic $s$ wave and the $d$ wave 
pairing, for a perfectly nested Fermi surface. 
Due to the nesting effect, coherence 
peaks are clearly visible in the charge (spin) DSF 
in the anisotropic $s$ wave ($d$ wave) case for any of the 
nesting vector $\qq=(\pm\pi,\pm\pi)$, while they are strongly suppressed 
in the $d$ wave (anisotropic $s$ wave) case (Fig.~\ref{fig:PhaseBZ}). 
The case of an isotropic Fermi surface is presented in Appendix~II\ref{app:isotropic}. 
Clearly the 
symmetry of the order parameter is reflected by the symmetry of the 
charge and spin DSF spectrum. 
Since the charge and spin DSF are complementary with respect to the spectral suppression of the sign reversing and sign preserving excitations, the phase sensitivity is enhanced when these two components are fully disentangled. 
This can be done by tuning the polarization dependence in the form factors
$W_{\bf e}^c$ and ${\bf W}_{\bf e}^s$ in Eq.~\eqref{eq:crosssectionb}. 

\paragraph{Origin of the phase sensitivity}
It occurs that 
a strong 
dependence on the SC order parameter phase in DSF is found 
for transferred momenta $q \gtrsim 0.1 k_F$ and when the transferred 
energy is close to twice the energy of the SC gap 
$\hbar \omega \simeq 2|\Delta|$. 
This sensitivity to the SC order parameter phase
can be better understood if we further confine our discussion 
to the case where not only the transferred momentum is rather 
large, i.e., $q \simeq k_F$, but also the momentum $\bf k$ is on the Fermi surface. 
The main contributions to the DSF correspond in fact to those 
excitations close to the Fermi surface which fulfill the 
conditions $\varepsilon_{\bf k} \ll \Delta_{\bf k}$ 
and $\varepsilon_{{\bf k} + {\bf q}} \ll \Delta_{{\bf k} + {\bf q}}$.
Assuming an unconventional superconductor with a pairing governed by a 
phase dependent order parameter 
$\Delta_{\bf k} =|\Delta_{\bf k}|e^{i\phi_{\bf k}}$ the 
DSF in Eq.~\eqref{StructureFactor} for excitations near the Fermi 
surface ($\hbar\omega \lesssim 2\Delta$)
becomes
\begin{align}\label{SA_FS}
\chi^{c, s} ({\bf q}, \hbar\omega)
 \approx \sum_{{\bf k}\in FS} & 
[
1 \pm \cos(\phi_\kk - \phi_{\kk+\qq}) 
]
\nonumber \\
& \times \delta(\hbar\omega-|\Delta_{\kk+\qq}| - |\Delta_\kk|).
\end{align}
Thus, the momentum-dependent intensity distribution of 
the low energy DSF mainly represents the variation of the SC order parameter phase along the 
Fermi surface. 

\paragraph{Comparison with other spectroscopies}
In principle also other two-particle spectroscopies 
(see, e.g., Ref.~\onlinecite{Ament2011} for an overview) can be \emph{directly} 
sensitive to the DSF of superconductors. Even if none can match RIXS in measuring both spin and charge DSF of superconductors, already probing either of the two is in general challenging as in, e.g., electron energy loss spectroscopy (EELS) one cannot reliably measure spectra with high momentum transfers, inelastic neutron scattering (INS) does not directly probe the charge DSF and non-resonant inelastic x-ray scattering (NIXS) is extremely photon-hungry. 
Nevertheless, transition amplitudes of the same type as in 
Eqs.~(\ref{StructureFactor},\ref{SA_FS}) are also encountered when determining the scattering rate of conduction 
electrons in the presence of impurities, as observed in the surface-sensitive 
STS~\cite{Hoffman2002,McElroy2003,Hanaguri2007,Kohsaka2008,Hanaguri2009,Hanke12,Sykora11}. 
This is because Eqs.~(\ref{StructureFactor},\ref{SA_FS}) have a similar 
structure as the ones which are known to govern the quasiparticle interference 
[in which case the transition amplitudes, whose sum over the 
momentum ${\bf k}$ contribute to the DSF in Eq.~(\ref{StructureFactor}), are termed 
`coherence factors'] 
in the presence of impurities.
Since the quasiparticle interference patterns explored by STS have turned out 
to be very successful in uncover the pairing symmetries of the unconventional 
SC~\cite{Hoffman2002,McElroy2003,Hanaguri2007,Kohsaka2008,Hanaguri2009,Hanke12},
this gauges the potential of RIXS to observe and unravel symmetries of SC pairing and pairing-mediators. 
Compared to STS, however, RIXS has a succinct conceptual advantage. 
Whereas the theoretical interpretation of STS in  
the framework of quasiparticle interference relies crucially on the form 
of the underlying impurity system showing various components 
of scattering~\cite{Sykora11}, in the case of RIXS 
the interpretation of spectroscopic features does neither rely on 
the presence of impurities in the superconductor nor on the modeling thereof.

\paragraph{Conclusions}
We have shown that RIXS, in contrast to other well-known two-particle spectroscopies, is directly sensitive to the spin \emph{and} to the charge dynamical structure factor (DSF) of a superconductor. 
In particular we have shown that the DSF of a superconductor observed in RIXS is very sensitive to the symmetry of the order parameter. This is rooted in the quasiparticle spectra reflecting sign-reversing excitations at large transferred momenta which arise for order parameters with a phase that varies over the Fermi surface. 
This, together with the recent experimental successes of RIXS, including in particular the major enhancements in resolution and pioneering study of hole doped cuprates~\cite{Tacon2011}, establishes the potential of RIXS 
as a versatile and practical spectroscopic technique to investigate the fundamental properties of superconducting materials.

\paragraph{Acknowledgments} 
The authors thank K.~W.~Becker for fruitful discussions related to this work. K.\,W. acknowledges support from the Alexander von Humboldt Foundation. This research benefited from the RIXS collaboration supported by the
Computational Materials Science Network (CMSN) program of the Division
of Materials Science and Engineering, U.S. Department of Energy, Grant
No. DE-FG02-08ER46540.


\quad\\
{\Large \bf Appendices}

\section{I - Dynamical structure factor for small  ($q\ll k_F$) scattering}
\label{app:smallq}

\begin{figure}[t]
\centering
\includegraphics[width=1\columnwidth]{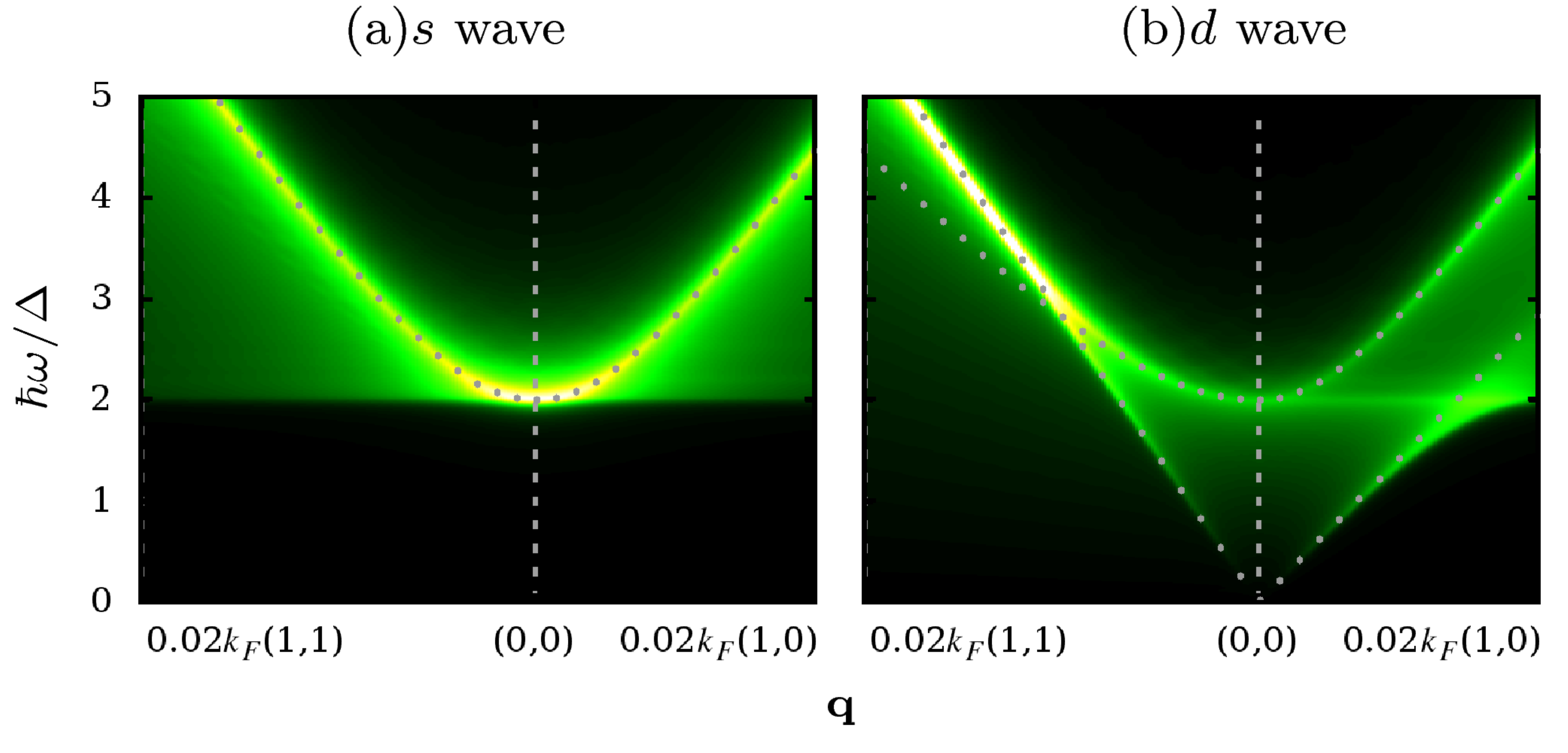}
\caption
{
Dynamical structure factor of Eq.~(3) 
in the main text of the paper
for an isotropic Fermi surface 
with parabolic dispersion as a function of transferred momentum $\qq$ near (0,0) 
for $s$ wave (a) and $d$ wave (b) superconductors. The 
highest energy dispersion is drawn (dotted line) for gapped 
(a) excitations in the conventional SC state, while in the case 
of an unconventional superconductor (b) energy dispersions 
(dotted lines) refer to excitations with momenta on nodal (gapless) 
and anti-nodal points (gapped) of the order parameter.
}
\label{fig:IsotropicFS}
\end{figure}

In what follows we discuss the effect of the momentum dependence 
of the SC gap magnitude on the RIXS spectrum in the case transferred 
momentum $q/ k_F\ll 1$. 
In this regime the spectral 
intensity is rather insensitive with respect to  points where the sign of the 
SC order parameter changes. 
In the normal state ($\Deltak=0$) the dynamical structure factor (DSF) is  
non-zero only for particle-hole excitations close to the Fermi surface 
leading to a coherence peak in the spectrum which disperses according to 
the Fermi velocity. In Fig.~\ref{fig:IsotropicFS} the DSF is shown for a conventional $s$ wave (a) and for 
an unconventional $d$ wave (b) superconductor, calculated using Eq. (3) 
in the main text of the paper for an isotropic Fermi surface as a function of the 
transferred momentum $q\ll k_F$ close to (0,0). Clearly seen is the absence 
of spectral weight for $\hbar \omega < 2\Delta$ in the conventional case (a), 
whereas at higher energies the coherence peak follows the dispersion 
$\varepsilon_{\bf k}$ of the bare electrons (dotted line).
Instead, for a $d$-wave superconductor (b) two types of excitations 
contribute to the low energy spectrum, related 
to two different regimes of momenta ${\bf k}$ contributing to the 
sum of Eq.~(3) in the main text of the paper. 
If the momentum ${\bf k}$ is 
close to the nodal lines, i.e., the Brillouin zone diagonals shown 
in Fig.~1 of the main text of the paper, where $\Delta_{\bf k} \approx 0$,
the excitations are particle-hole like with an energy $\propto q$.

The second type of excitations is provided by momenta close to the 
anti-nodal points ${\bf k}_A$ where $\Delta_{{\bf k}_A} = \Delta$ 
is the maximum gap value. Here, for small values of ${\bf q}$ the 
transition energy $E_{{\bf k}_A + {\bf q}} + E_{{\bf k}_A}$ is 
approximately
\begin{equation}\label{GappedDispersion}
E_{{\bf k}_A + {\bf q}} + E_{{\bf k}_A} \approx 
2\sqrt{|\Delta|^2+\left(\frac{\hbar {\bf v}_{{\bf k}_A}\cdot\qq}{2}\right)^2} \ ,
\end{equation}
where $\vvk=\rm{d}\epsk/\hbar \rm{d}\kk$ is the electron group 
velocity. Thus, as it is seen in Fig.~\ref{fig:IsotropicFS}(b) for 
$E>2\Delta$ we obtain a coherence peak showing the known dispersion of 
Fig.~\ref{fig:IsotropicFS}(a), but with less spectral weight since the 
excitations are restricted to the vicinity of the anti-nodal points. 
The ${\bf q}$-dependence of the excitation energy given by 
Eq.~\eqref{GappedDispersion} determines  the behavior of the coherence 
peak in a typical RIXS spectrum in the SC state. An additional
coherent excitation showing a linear dispersion down to $\omega=0$ 
for ${\bf q} \rightarrow 0$ indicates the presence of nodes in 
the SC order parameter.

\section{II - Dynamical structure factor for an isotropic Fermi surface superconductor}
\label{app:isotropic}

\begin{figure}
\centering
\includegraphics[width=1\columnwidth]{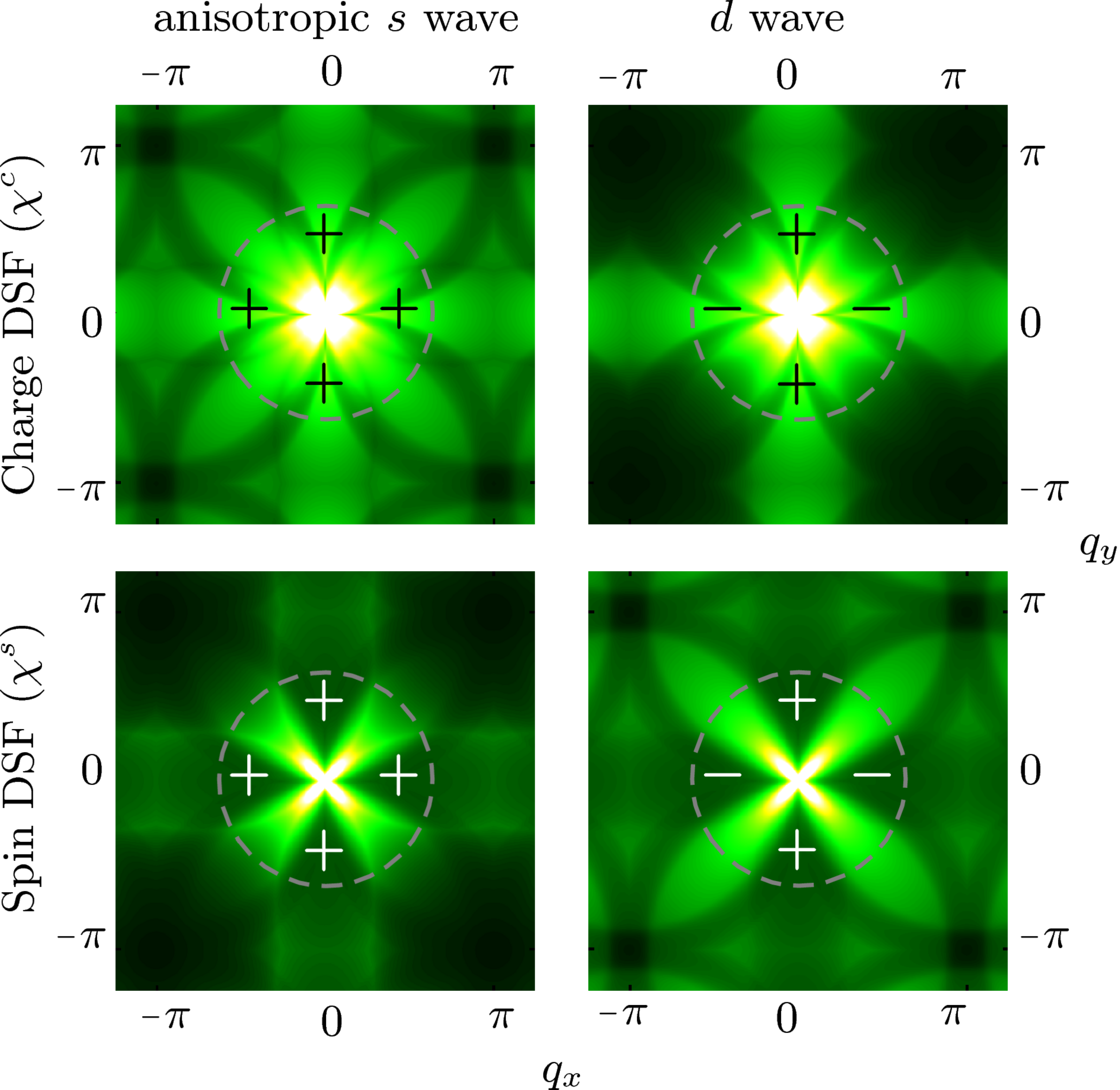}
\caption
{
Charge ($\chi^c$) and spin ($\chi^s$) DSF at fixed energy 
$\hbar \omega=2|\Delta|$, as a function of transferred momentum
for an anisotropic $s$ wave 
and a $d$ wave SC with an isotropic Fermi surface.
}
\label{fig:PhaseBZiso}
\end{figure}

The spin and charge DSF spectra are shown in Fig.~\ref{fig:PhaseBZiso}, for an isotropic Fermi surface and for different symmetries of the order parameter.
As one can see, $d$ wave (anisotropic $s$ wave) pairing strongly suppresses the 
spectral weight in the charge (spin) DSF for transferred momenta that correspond to sign reversing (sign preserving) 
excitations, while no suppression occurs in the anisotropic $s$ wave ($d$ wave) case. 

\section{III - Dynamical structure factor for strongly correlated electron system}
\label{app:strong}

\begin{figure*}[t]
\centering
\includegraphics[width=1.8\columnwidth]{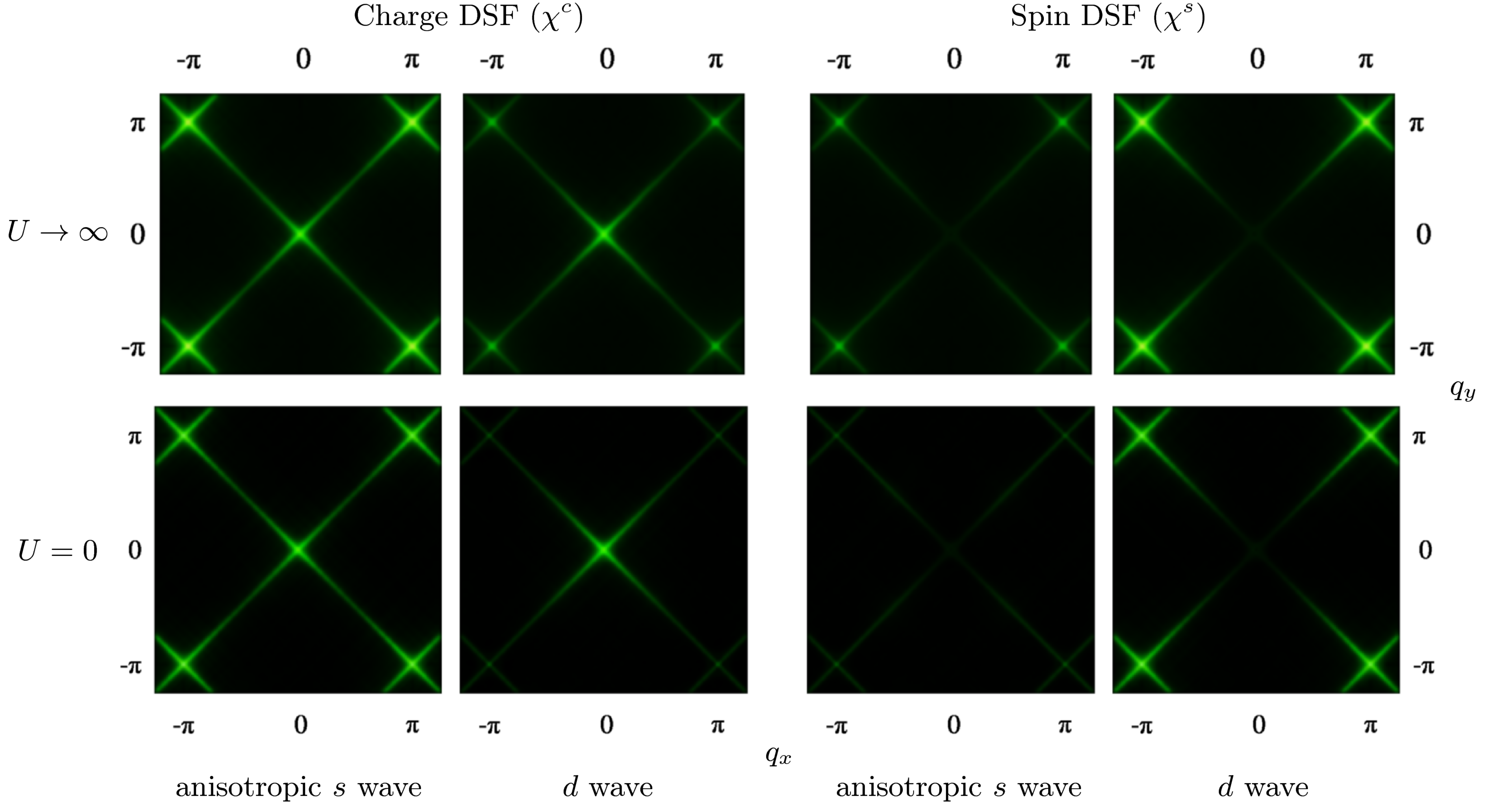}
\caption
{
Charge ($\chi^c$) and spin ($\chi^s$) DSF at the quasiparticle gap level, as a function of transferred momentum for an anisotropic $s$ wave (first and third columns) and a $d$ wave (second and fourth columns) superconductor, for a cuprate-like system with a nested Fermi surface, with an on-site repulsion $U\rightarrow\infty$ (first row) and $U=0$ (second row). In the infinite repulsion case, the Gutzwiller approximation is used, assuming $D=\langle{\mathcal D}_{\sigma}(i)\rangle=0.8$ (hole doped system, cf. main text), and the gap level is renormalized as $2|\Delta|\rightarrow 2D|\Delta|$. 
In the $d$ wave (anisotropic $s$ wave) superconductor, the charge (spin) DSF peak at $\qq=(\pi,\pi)$ is suppressed, for both the infinitely strongly correlated and the uncorrelated case.
}
\label{fig:U}
\end{figure*}

The main aim of this part of the appendix
is to show that our main result for the phase dependence
of the dynamical structure factor (DSF) holds also in the 
presence of strong Coulomb repulsion of the conduction 
electrons. To simplify the presentation of our arguments 
we will focus on
the limit of infinite strong correlations
$U\rightarrow \infty$ where double occupancy is forbidden~\cite{Trugman1990,Becca2000}. As is
well-known using a Schrieffer-Wolf transformation the Hubbard model
can then be replaced with a $t$-$J$ model where the superexchange $J\rightarrow 0$, cf. \cite{Chao1978}.
Such regime of very strong correlations may be still
regarded as rather realistic in describing many basic properties
of the strongly transition metal oxides such as cuprates
in the overdoped limit~\cite{Putikka1992,Tandon1999,Cosentini1998}.
The superconductor is well described by the following Hamiltonian
\begin{eqnarray}
\label{sc1}
 \mathcal H &=& \sum_{{\mathbf k}\sigma} \varepsilon_{\mathbf k}
  \hat{d}_{{\mathbf k} \sigma}^\dagger  \hat{d}_{{\mathbf k} \sigma} 
  -  \sum_{\mathbf k} 
   \Delta_{\mathbf k} 
   \hat{d}_{{\mathbf k} \uparrow}^\dagger  \hat{d}_{-{\mathbf k} \downarrow}^\dagger 
   + \mbox{h.c.} 
\ .
\end{eqnarray}
The operators 
$ \hat {d}_{i\sigma}^\dagger =  {d}_{i\sigma}^\dagger(1- n_{i,-\sigma})$,  and
$\hat {d}_{i\sigma} =  {d}_{i\sigma}(1- n_{i,-\sigma})$ 
are Hubbard creation and annihilation operators which enter since 
doubly occupancies of local sites are strictly forbidden due to the presence of strong 
electronic correlations. They obey unusual anticommutator relations. For instance,  
$[\hat {d}_{i\sigma}, \hat {d}_{j\sigma}^\dag   ]_+ = \delta_{ij}{\mathcal D}_{\sigma}(i)$, with
${\mathcal D}_{\sigma}(i)= 1 - n_{i, -\sigma}$.

For simplicity let us consider the spin DSF
$\chi^{s}({\bf q},\omega)$ which is defined by the
following correlation function
\begin{eqnarray} 
 \label{sc2}
   \chi^{s}({\bf q},\omega)  &=&
  i \int_0^\infty dt \, \langle S_{-{\mathbf q}}^z(t) 
 S_{\mathbf q}^z \rangle_{\mathcal H} \ e^{i(\omega +i \eta)t}   .
\end{eqnarray}
Here, the time dependence and the expectation 
value are formed with  Hamiltonian \eqref{sc1}. To calculate the expectation value and 
the dynamical behavior, we diagonalize the Hamiltonian using new 
approximate quasiparticle operators $\hat{\gamma}_{{\bf k},\sigma}$, which are
related to the original correlated $d$ electron operators via 
$\hat{d}_{\kk\uparrow}=\uk^* \hat{\gamma}_{\kk\uparrow}-\vk 
\hat{\gamma}^\dag_{-\kk\downarrow}$ and $\hat{d}_{\kk\downarrow}=\uk^* 
\hat{\gamma}_{\kk\downarrow}+\vk \hat{\gamma}^\dag_{-\kk\uparrow}$. In the 
case of a sufficiently large hole concentration the operator ${\mathcal D}_{\sigma}(i)$
can approximately be replaced by its expectation value $D$ and the
Bogoliubov quasiparticle operators fulfill the following relations~\cite{Sykora2009}:  $[\mathcal H,
  \hat{\gamma}_{{\bf k},\sigma}^\dag] = E_{\bf k} \hat{\gamma}_{{\bf k},\sigma}^\dag$, where 
$E_{\bf k} = \sqrt{ \varepsilon_{\bf k}^2 + D^2 \Delta_{\bf k}^2}$ and $D = 1 - n/2$.
Replacing all $\hat{d}_{\bf k\sigma}^{(\dagger)}$ operators by  
the quasiparticle operators $\hat{\gamma}_{\bf k\sigma}^{(\dagger)}$, the time dependence  
can easily be evaluated. 
Each of the remaining expectation values contains a product of four quasiparticle operators. 
A final factorization leads to the following expectation values with two quasiparticle operators
\begin{eqnarray}
\label{sc3}
\langle \hat{\gamma}_{{\bf k},\sigma}^\dag \hat{\gamma}_{{\bf k},\sigma} \rangle &=&
\frac{1}{2}\left(1 + \frac{\varepsilon_{\bf k}}{E_{\mathbf k}}  \right) n_{\bf k} +
\frac{1}{2}\left(1 - \frac{\varepsilon_{\bf k}}{E_{\mathbf k}}  \right) m_{\bf k} + \nonumber \\
&-& \frac{D^3 | \Delta_{\bf k}|^2}{2 E_{\bf k}^2}, \nonumber \\
\langle \hat{\gamma}_{{\bf k},\sigma} \hat{\gamma}_{{\bf k},\sigma}^\dag \rangle &=&
D - \langle \hat{\gamma}_{{\bf k},\sigma}^\dag \hat{\gamma}_{{\bf k},\sigma} \rangle,
\end{eqnarray}
where $D= \langle {\mathcal D_\sigma (i)} \rangle = 1 -n/2$ and $n_{\bf k}$ and $m_{\bf k}$ are defined by
 $n_{\bf k}= \langle \hat d_{{\bf k}\sigma}^\dagger 
\hat d_{{\bf k}\sigma}\rangle$ 
and  $m_{\bf k}= \langle \hat d_{{\bf k}\sigma} \hat d_{{\bf k}\sigma}^\dagger\rangle$. 
They are evaluated using the Gutzwiller approximation (cf. Ref.~\onlinecite{Fazekas}), as
$n_{{\bf k}} = (D-q)  + q  \, f( \varepsilon_{\bf k})$ and $m_{\bf k} = D - n_{{\bf k}}$, 
where $q = (1 - n) / (1-n/2)$ and $f( \varepsilon_{\bf k})$ is the 
Fermi function at $T=0$. 

Finally, one obtains the spin and charge DSF in the strongly correlated case
 \begin{eqnarray}
\label{sc4}
 && \chi^{c,s}({\bf q},\omega) = 
\nonumber \\
 &&
\sum_{\bf k} \left[ \frac{A^\pm({\bf k}, {\bf q})}{E_{\bf k}+E_{{\bf k}+{\bf q}}- (\omega + i \eta)} 
 \langle \hat{\gamma}_{{\bf k}+{\bf q},\downarrow} \hat{\gamma}_{{\bf k}+{\bf q},\downarrow}^\dag \rangle
 \langle \hat{\gamma}_{{\bf k},\uparrow} \hat{\gamma}_{{\bf k},\uparrow}^\dag \rangle
+ \nonumber 
 \right.
  \\
 && 
 \left. 
 + \frac{A^\pm({\bf k}, {\bf q})}{-E_{\bf k}- E_{{\bf k}+{\bf q}}- (\omega + i \eta)} 
 \langle \hat{\gamma}_{{\bf k}+{\bf q},\uparrow}^\dag \hat{\gamma}_{{\bf k}+{\bf q},\uparrow} \rangle
 \langle \hat{\gamma}_{{\bf k},\downarrow}^\dag \hat{\gamma}_{{\bf k},\downarrow} \rangle
+ \nonumber 
 \right.
  \\
&&
\left.
+ \frac{2 A^\mp({\bf k}, {\bf q})}{E_{\bf k}- E_{{\bf k}+{\bf q}}- (\omega + i \eta)} 
 \langle \hat{\gamma}_{{\bf k}+{\bf q},\uparrow}^\dag \hat{\gamma}_{{\bf k}+{\bf q},\uparrow} \rangle
 \langle \hat{\gamma}_{{\bf k},\uparrow} \hat{\gamma}_{{\bf k},\uparrow}^\dag \rangle \nonumber  
 \right],
\\&&
\end{eqnarray}
where the coherence factors $A^+({\bf k}, {\bf q})$ and $A^-({\bf k}, {\bf q})$ are defined by 
\begin{eqnarray}
\label{sc5}
A^\pm({\bf k}, {\bf q}) &=&
1 \pm \frac{ D^2 \mbox{Re}(\Delta_{\bf k} \Delta_{{\bf k}+{\bf q}}^*)  \mp \varepsilon_{\bf k} \varepsilon_{{\bf k}+{\bf q}}
}
{E_{\bf k} E_{{\bf k}+{\bf q}}  }
\end{eqnarray}

To summarize, the main effect of correlations in the $U\rightarrow\infty$ case is to rescale the magnitude of the order parameter. 
In fact, up to a renormalized gap function $\Delta_\kk\rightarrow D\Delta_\kk$, 
the coherence factors in Eq.~\eqref{sc5} have the same form as in the uncorrelated case [cf. Eq.~(3) of the letter]. 
Due to this renormalization, the quasiparticle excitations gap is lowered in energy by a factor $D$. 
Moreover, the phase sensitivity of the DSF is reduced by the presence of the third term of Eq.~\eqref{sc4}. 
However, when the hole doping is rather large ($D\approx1$, $n\ll1$), 
the order parameter decrease is negligible, while the main contributions to the quasiparticle spectrum are given by the first term in Eq.~\eqref{sc4}. 
As a consequence, the phase sensitivity of the DSF is not affected. 
In Fig.~\ref{fig:U} we compare the spin and charge DSF of a strongly correlated hole doped system with those of an uncorrelated one (cf. Fig.~3 of the letter), for different order parameter symmetries ($d$ wave and anisotropic $s$ wave). 
As one can see, the presence of electronic correlations does not change RIXS spectra qualitatively.
Hence the dynamical structure factor in a strongly correlated electron system is governed, as well as in an uncorrelated one, by coherence factors which are responsible for the sensitivity of RIXS spectra to the order parameter phase.


\begin{thebibliography}{10}%
\makeatletter
\providecommand \@ifxundefined [1]{%
 \ifx #1\undefined \expandafter \@firstoftwo
 \else \expandafter \@secondoftwo
\fi
}%
\providecommand \@ifnum [1]{%
 \ifnum #1\expandafter \@firstoftwo
 \else \expandafter \@secondoftwo
\fi
}%
\providecommand \enquote [1]{``#1''}%
\providecommand \bibnamefont  [1]{#1}%
\providecommand \bibfnamefont [1]{#1}%
\providecommand \citenamefont [1]{#1}%
\providecommand\href[0]{\@sanitize\@href}%
\providecommand\@href[1]{\endgroup\@@startlink{#1}\endgroup\@@href}%
\providecommand\@@href[1]{#1\@@endlink}%
\providecommand \@sanitize [0]{\begingroup\catcode`\&12\catcode`\#12\relax}%
\@ifxundefined \pdfoutput {\@firstoftwo}{%
 \@ifnum{\z@=\pdfoutput}{\@firstoftwo}{\@secondoftwo}%
}{%
 \providecommand\@@startlink[1]{\leavevmode}%
 \providecommand\@@endlink[0]{}%
}{%
 \providecommand\@@startlink[1]{%
  \leavevmode
  \pdfstartlink
   attr{/Border[0 0 1 ]/H/I/C[0 1 1]}%
   user{/Subtype/Link/A<</Type/Action/S/URI/URI(#1)>>}%
  \relax
 }%
 \providecommand\@@endlink[0]{\pdfendlink}%
}%
\providecommand \url  [0]{\begingroup\@sanitize \@url }%
\providecommand \@url [1]{\endgroup\@href {#1}{\urlprefix}}%
\providecommand \urlprefix [0]{URL }%
\providecommand \Eprint[0]{\href }%
\@ifxundefined \urlstyle {%
  \providecommand \doi [1]{doi:\discretionary{}{}{}#1}%
}{%
  \providecommand \doi [0]{doi:\discretionary{}{}{}\begingroup
  \urlstyle{rm}\Url }%
}%
\providecommand \doibase [0]{http://dx.doi.org/}%
\providecommand \Doi[1]{\href{\doibase#1}}%
\providecommand \bibAnnote [3]{%
  \BibitemShut{#1}%
  \begin{quotation}\noindent
    \textsc{Key:}\ #2\\\textsc{Annotation:}\ #3%
  \end{quotation}%
}%
\providecommand \bibAnnoteFile [2]{%
  \IfFileExists{#2}{\bibAnnote {#1} {#2} {\input{#2}}}{}%
}%
\providecommand \typeout [0]{\immediate \write \m@ne }%
\providecommand \selectlanguage [0]{\@gobble}%
\providecommand \bibinfo [0]{\@secondoftwo}%
\providecommand \bibfield [0]{\@secondoftwo}%
\providecommand \translation [1]{[#1]}%
\providecommand \BibitemOpen[0]{}%
\providecommand \bibitemStop [0]{}%
\providecommand \bibitemNoStop [0]{.\EOS\space}%
\providecommand \EOS [0]{\spacefactor3000\relax}%
\providecommand \BibitemShut [1]{\csname bibitem#1\endcsname}%
\bibitem{Schuelke2007}%
  \BibitemOpen
  \bibfield{author}{%
  \bibinfo {author} {\bibfnamefont{W.}~\bibnamefont{Sch{\"u}lke}},\ }%
  \emph{\bibinfo {title} {Electron Dynamics by Inelastic {X}-Ray Scattering}}\
  (\bibinfo {publisher} {Oxford University Press},\ \bibinfo {address}
  {Oxford},\ \bibinfo {year} {2007})%
  \bibAnnoteFile{NoStop}{Schuelke2007}%
\bibitem{Ament2011}%
  \BibitemOpen
  \bibfield{author}{%
  \bibinfo {author} {\bibfnamefont{L.~J.~P.}\ \bibnamefont{Ament}}, \bibinfo
  {author} {\bibfnamefont{M.}~\bibnamefont{van Veenendaal}}, \bibinfo {author}
  {\bibfnamefont{T.~P.}\ \bibnamefont{Devereaux}}, \bibinfo {author}
  {\bibfnamefont{J.~P.}\ \bibnamefont{Hill}},\ and\ \bibinfo {author}
  {\bibfnamefont{J.}~\bibnamefont{van~den Brink}},\ }%
  \bibfield{journal}{%
  \bibinfo {journal} {Rev. Mod. Phys.}\ }%
  \textbf{\bibinfo {volume} {83}},\ \bibinfo {pages} {705} (\bibinfo {year}
  {2011})%
  \bibAnnoteFile{NoStop}{Ament2011}%
\bibitem{Braicovich2009a}%
  \BibitemOpen
  \bibfield{author}{%
  \bibinfo {author} {\bibfnamefont{L.}~\bibnamefont{Braicovich}}, \bibinfo
  {author} {\bibfnamefont{L.~J.~P.}\ \bibnamefont{Ament}}, \bibinfo {author}
  {\bibfnamefont{V.}~\bibnamefont{Bisogni}}, \bibinfo {author}
  {\bibfnamefont{F.}~\bibnamefont{Forte}}, \bibinfo {author}
  {\bibfnamefont{C.}~\bibnamefont{Aruta}}, \bibinfo {author}
  {\bibfnamefont{G.}~\bibnamefont{Balestrino}}, \bibinfo {author}
  {\bibfnamefont{N.~B.}\ \bibnamefont{Brookes}}, \bibinfo {author}
  {\bibfnamefont{G.~M.}\ \bibnamefont{De~Luca}}, \bibinfo {author}
  {\bibfnamefont{P.~G.}\ \bibnamefont{Medaglia}}, \bibinfo {author}
  {\bibfnamefont{F.}~\bibnamefont{Miletto~Granozio}}, \bibinfo {author}
  {\bibfnamefont{M.}~\bibnamefont{Radovic}}, \bibinfo {author}
  {\bibfnamefont{M.}~\bibnamefont{Salluzzo}}, \bibinfo {author}
  {\bibfnamefont{J.}~\bibnamefont{van~den Brink}},\ and\ \bibinfo {author}
  {\bibfnamefont{G.}~\bibnamefont{Ghiringhelli}},\ }%
  \bibfield{journal}{%
  \bibinfo {journal} {Phys. Rev. Lett.}\ }%
  \textbf{\bibinfo {volume} {102}},\ \bibinfo {pages} {167401} (\bibinfo {year}
  {2009})%
  \bibAnnoteFile{NoStop}{Braicovich2009a}%
\bibitem{Braicovich2009b}%
  \BibitemOpen
  \bibfield{author}{%
  \bibinfo {author} {\bibfnamefont{L.}~\bibnamefont{Braicovich}}, \bibinfo
  {author} {\bibfnamefont{J.}~\bibnamefont{van~den Brink}}, \bibinfo {author}
  {\bibfnamefont{V.}~\bibnamefont{Bisogni}}, \bibinfo {author}
  {\bibfnamefont{M.}~\bibnamefont{Moretti~Sala}}, \bibinfo {author}
  {\bibfnamefont{L.~J.~P.}\ \bibnamefont{Ament}}, \bibinfo {author}
  {\bibfnamefont{N.~B.}\ \bibnamefont{Brookes}}, \bibinfo {author}
  {\bibfnamefont{G.~M.}\ \bibnamefont{De~Luca}}, \bibinfo {author}
  {\bibfnamefont{M.}~\bibnamefont{Salluzzo}}, \bibinfo {author}
  {\bibfnamefont{T.}~\bibnamefont{Schmitt}}, \bibinfo {author}
  {\bibfnamefont{V.~N.}\ \bibnamefont{Strocov}},\ and\ \bibinfo {author}
  {\bibfnamefont{G.}~\bibnamefont{Ghiringhelli}},\ }%
  \bibfield{journal}{%
  \bibinfo {journal} {Phys. Rev. Lett.}\ }%
  \textbf{\bibinfo {volume} {104}},\ \bibinfo {pages} {077002} (\bibinfo {year}
  {2010})%
  \bibAnnoteFile{NoStop}{Braicovich2009b}%
\bibitem{Hill2008}%
  \BibitemOpen
  \bibfield{author}{%
  \bibinfo {author} {\bibfnamefont{J.~P.}\ \bibnamefont{Hill}}, \bibinfo
  {author} {\bibfnamefont{G.}~\bibnamefont{Blumberg}}, \bibinfo {author}
  {\bibfnamefont{Y.-J.}\ \bibnamefont{Kim}}, \bibinfo {author}
  {\bibfnamefont{D.~S.}\ \bibnamefont{Ellis}}, \bibinfo {author}
  {\bibfnamefont{S.}~\bibnamefont{Wakimoto}}, \bibinfo {author}
  {\bibfnamefont{R.~J.}\ \bibnamefont{Birgeneau}}, \bibinfo {author}
  {\bibfnamefont{S.}~\bibnamefont{Komiya}}, \bibinfo {author}
  {\bibfnamefont{Y.}~\bibnamefont{Ando}}, \bibinfo {author}
  {\bibfnamefont{B.}~\bibnamefont{Liang}}, \bibinfo {author}
  {\bibfnamefont{R.~L.}\ \bibnamefont{Greene}}, \bibinfo {author}
  {\bibfnamefont{D.}~\bibnamefont{Casa}},\ and\ \bibinfo {author}
  {\bibfnamefont{T.}~\bibnamefont{Gog}},\ }%
  \bibfield{journal}{%
  \bibinfo {journal} {Phys. Rev. Lett.}\ }%
  \textbf{\bibinfo {volume} {100}},\ \bibinfo {pages} {097001} (\bibinfo {year}
  {2008})%
  \bibAnnoteFile{NoStop}{Hill2008}%
\bibitem{Tacon2011}%
  \BibitemOpen
  \bibfield{author}{%
  \bibinfo {author} {\bibfnamefont{M.}~\bibnamefont{Le~Tacon}}, \bibinfo
  {author} {\bibfnamefont{G.}~\bibnamefont{Ghiringhelli}}, \bibinfo {author}
  {\bibfnamefont{J.}~\bibnamefont{Chaloupka}}, \bibinfo {author}
  {\bibfnamefont{M.~M.}\ \bibnamefont{Sala}}, \bibinfo {author}
  {\bibfnamefont{V.}~\bibnamefont{Hinkov}}, \bibinfo {author}
  {\bibfnamefont{M.~W.}\ \bibnamefont{Haverkort}}, \bibinfo {author}
  {\bibfnamefont{M.}~\bibnamefont{Minola}}, \bibinfo {author}
  {\bibfnamefont{M.}~\bibnamefont{Bakr}}, \bibinfo {author}
  {\bibfnamefont{K.~J.}\ \bibnamefont{Zhou}}, \bibinfo {author}
  {\bibfnamefont{S.}~\bibnamefont{Blanco-Canosa}}, \bibinfo {author}
  {\bibfnamefont{C.}~\bibnamefont{Monney}}, \bibinfo {author}
  {\bibfnamefont{Y.~T.}\ \bibnamefont{Song}}, \bibinfo {author}
  {\bibfnamefont{G.~L.}\ \bibnamefont{Sun}}, \bibinfo {author}
  {\bibfnamefont{C.~T.}\ \bibnamefont{Lin}}, \bibinfo {author}
  {\bibfnamefont{G.~M.}\ \bibnamefont{De~Luca}}, \bibinfo {author}
  {\bibfnamefont{M.}~\bibnamefont{Salluzzo}}, \bibinfo {author}
  {\bibfnamefont{G.}~\bibnamefont{Khaliullin}}, \bibinfo {author}
  {\bibfnamefont{T.}~\bibnamefont{Schmitt}}, \bibinfo {author}
  {\bibfnamefont{L.}~\bibnamefont{Braicovich}},\ and\ \bibinfo {author}
  {\bibfnamefont{B.}~\bibnamefont{Keimer}},\ }%
  \bibfield{journal}{%
  \bibinfo {journal} {{Nature Physics}}\ }%
  \textbf{\bibinfo {volume} {{7}}},\ \bibinfo {pages} {725} (\bibinfo {year}
  {{2011}})%
  \bibAnnoteFile{NoStop}{Tacon2011}%
\bibitem{Kim2012b}%
  \BibitemOpen
  \bibfield{author}{%
  \bibinfo {author} {\bibfnamefont{J.}~\bibnamefont{Kim}}, \bibinfo {author}
  {\bibfnamefont{A.~H.}\ \bibnamefont{Said}}, \bibinfo {author}
  {\bibfnamefont{D.}~\bibnamefont{Casa}}, \bibinfo {author}
  {\bibfnamefont{M.~H.}\ \bibnamefont{Upton}}, \bibinfo {author}
  {\bibfnamefont{T.}~\bibnamefont{Gog}}, \bibinfo {author}
  {\bibfnamefont{M.}~\bibnamefont{Daghofer}}, \bibinfo {author}
  {\bibfnamefont{G.}~\bibnamefont{Jackeli}}, \bibinfo {author}
  {\bibfnamefont{J.}~\bibnamefont{van~den Brink}}, \bibinfo {author}
  {\bibfnamefont{G.}~\bibnamefont{Khaliullin}},\ and\ \bibinfo {author}
  {\bibfnamefont{B.~J.}\ \bibnamefont{Kim}},\ }%
  \bibfield{journal}{%
  \bibinfo {journal} {Arxiv},\ \bibinfo {pages} {1205.5337}}%
   (\bibinfo {year} {2012})%
  \bibAnnoteFile{NoStop}{Kim2012b}%
\bibitem{Dean2012}%
  \BibitemOpen
  \bibfield{author}{%
  \bibinfo {author} {\bibfnamefont{M.~P.~M.}\ \bibnamefont{Dean}}, \bibinfo
  {author} {\bibfnamefont{R.~S.}\ \bibnamefont{Springell}}, \bibinfo {author}
  {\bibfnamefont{C.}~\bibnamefont{Monney}}, \bibinfo {author}
  {\bibfnamefont{K.~J.}\ \bibnamefont{Zhou}}, \bibinfo {author}
  {\bibfnamefont{J.}~\bibnamefont{Pereiro}}, \bibinfo {author}
  {\bibfnamefont{I.}~\bibnamefont{Bo\v{z}ovi\'{c}}}, \bibinfo {author}
  {\bibfnamefont{B.}~\bibnamefont{Dalla~Piazza}}, \bibinfo {author}
  {\bibfnamefont{H.~M.}\ \bibnamefont{R\o{}nnow}}, \bibinfo {author}
  {\bibfnamefont{E.}~\bibnamefont{Morenzoni}}, \bibinfo {author}
  {\bibfnamefont{J.}~\bibnamefont{van~den Brink}}, \bibinfo {author}
  {\bibfnamefont{T.}~\bibnamefont{Schmitt}},\ and\ \bibinfo {author}
  {\bibfnamefont{J.~P.}\ \bibnamefont{Hill}},\ }%
  \bibfield{journal}{%
  \bibinfo {journal} {Nature Materials}\ }%
  \textbf{\bibinfo {volume} {11}},\ \bibinfo {pages} {850} (\bibinfo {year}
  {2012})%
  \bibAnnoteFile{NoStop}{Dean2012}%
\bibitem{Ulrich2009}%
  \BibitemOpen
  \bibfield{author}{%
  \bibinfo {author} {\bibfnamefont{C.}~\bibnamefont{Ulrich}}, \bibinfo {author}
  {\bibfnamefont{L.~J.~P.}\ \bibnamefont{Ament}}, \bibinfo {author}
  {\bibfnamefont{G.}~\bibnamefont{Ghiringhelli}}, \bibinfo {author}
  {\bibfnamefont{L.}~\bibnamefont{Braicovich}}, \bibinfo {author}
  {\bibfnamefont{M.}~\bibnamefont{Moretti~Sala}}, \bibinfo {author}
  {\bibfnamefont{N.}~\bibnamefont{Pezzotta}}, \bibinfo {author}
  {\bibfnamefont{T.}~\bibnamefont{Schmitt}}, \bibinfo {author}
  {\bibfnamefont{G.}~\bibnamefont{Khaliullin}}, \bibinfo {author}
  {\bibfnamefont{J.}~\bibnamefont{van~den Brink}}, \bibinfo {author}
  {\bibfnamefont{H.}~\bibnamefont{Roth}}, \bibinfo {author}
  {\bibfnamefont{T.}~\bibnamefont{Lorenz}},\ and\ \bibinfo {author}
  {\bibfnamefont{B.}~\bibnamefont{Keimer}},\ }%
  \bibfield{journal}{%
  \bibinfo {journal} {Phys. Rev. Lett.}\ }%
  \textbf{\bibinfo {volume} {103}},\ \bibinfo {pages} {107205} (\bibinfo {year}
  {2009})%
  \bibAnnoteFile{NoStop}{Ulrich2009}%
\bibitem{Schlappa2012}%
  \BibitemOpen
  \bibfield{author}{%
  \bibinfo {author} {\bibfnamefont{J.}~\bibnamefont{Schlappa}}, \bibinfo
  {author} {\bibfnamefont{K.}~\bibnamefont{Wohlfeld}}, \bibinfo {author}
  {\bibfnamefont{K.~J.}\ \bibnamefont{Zhou}}, \bibinfo {author}
  {\bibfnamefont{M.}~\bibnamefont{Mourigal}}, \bibinfo {author}
  {\bibfnamefont{M.~W.}\ \bibnamefont{Haverkort}}, \bibinfo {author}
  {\bibfnamefont{V.~N.}\ \bibnamefont{Strocov}}, \bibinfo {author}
  {\bibfnamefont{L.}~\bibnamefont{Hozoi}}, \bibinfo {author}
  {\bibfnamefont{C.}~\bibnamefont{Monney}}, \bibinfo {author}
  {\bibfnamefont{S.}~\bibnamefont{Nishimoto}}, \bibinfo {author}
  {\bibfnamefont{S.}~\bibnamefont{Singh}}, \bibinfo {author}
  {\bibfnamefont{A.}~\bibnamefont{Revcolevschi}}, \bibinfo {author}
  {\bibfnamefont{J.}~\bibnamefont{Caux}}, \bibinfo {author}
  {\bibfnamefont{L.}~\bibnamefont{Patthey}}, \bibinfo {author}
  {\bibfnamefont{H.~M.}\ \bibnamefont{Ronnow}}, \bibinfo {author}
  {\bibfnamefont{J.}~\bibnamefont{{van den Brink}}},\ and\ \bibinfo {author}
  {\bibfnamefont{T.}~\bibnamefont{Schmitt}},\ }%
  \bibfield{journal}{%
  \bibinfo {journal} {Nature}\ }%
  \textbf{\bibinfo {volume} {485}},\ \bibinfo {pages} {82} (\bibinfo {year}
  {2012})%
  \bibAnnoteFile{NoStop}{Schlappa2012}%
\bibitem{Yavas2010}%
  \BibitemOpen
  \bibfield{author}{%
  \bibinfo {author} {\bibfnamefont{H.}~\bibnamefont{Yavas}}, \bibinfo {author}
  {\bibfnamefont{M.}~\bibnamefont{van Veenendaal}}, \bibinfo {author}
  {\bibfnamefont{J.}~\bibnamefont{van~den Brink}}, \bibinfo {author}
  {\bibfnamefont{L.~J.~P.}\ \bibnamefont{Ament}}, \bibinfo {author}
  {\bibfnamefont{A.}~\bibnamefont{Alatas}}, \bibinfo {author}
  {\bibfnamefont{B.~M.}\ \bibnamefont{Leu}}, \bibinfo {author}
  {\bibfnamefont{M.-O.}\ \bibnamefont{Apostu}}, \bibinfo {author}
  {\bibfnamefont{N.}~\bibnamefont{Wizent}}, \bibinfo {author}
  {\bibfnamefont{G.}~\bibnamefont{Behr}}, \bibinfo {author}
  {\bibfnamefont{W.}~\bibnamefont{Sturhahn}}, \bibinfo {author}
  {\bibfnamefont{H.}~\bibnamefont{Sinn}},\ and\ \bibinfo {author}
  {\bibfnamefont{E.~E.}\ \bibnamefont{Alp}},\ }%
  \bibfield{journal}{%
  \bibinfo {journal} {J. Phys. Cond. Mat.}\ }%
  \textbf{\bibinfo {volume} {{22}}},\ \bibinfo {pages} {485601} (\bibinfo
  {year} {{2010}})%
  \bibAnnoteFile{NoStop}{Yavas2010}%
\bibitem{Groot1998}%
  \BibitemOpen
  \bibfield{author}{%
  \bibinfo {author} {\bibfnamefont{F.~M.~F.}\ \bibnamefont{de~Groot}}, \bibinfo
  {author} {\bibfnamefont{P.}~\bibnamefont{Kuiper}},\ and\ \bibinfo {author}
  {\bibfnamefont{G.~A.}\ \bibnamefont{Sawatzky}},\ }%
  \bibfield{journal}{%
  \bibinfo {journal} {Phys. Rev. B}\ }%
  \textbf{\bibinfo {volume} {57}},\ \bibinfo {pages} {14584} (\bibinfo {year}
  {1998})%
  \bibAnnoteFile{NoStop}{Groot1998}%
\bibitem{Ament2009a}%
  \BibitemOpen
  \bibfield{author}{%
  \bibinfo {author} {\bibfnamefont{L.~J.~P.}\ \bibnamefont{Ament}}, \bibinfo
  {author} {\bibfnamefont{G.}~\bibnamefont{Ghiringhelli}}, \bibinfo {author}
  {\bibfnamefont{M.}~\bibnamefont{Moretti~Sala}}, \bibinfo {author}
  {\bibfnamefont{L.}~\bibnamefont{Braicovich}},\ and\ \bibinfo {author}
  {\bibfnamefont{J.}~\bibnamefont{van~den Brink}},\ }%
  \bibfield{journal}{%
  \bibinfo {journal} {Phys. Rev. Lett.}\ }%
  \textbf{\bibinfo {volume} {103}},\ \bibinfo {pages} {117003} (\bibinfo {year}
  {2009})%
  \bibAnnoteFile{NoStop}{Ament2009a}%
\bibitem{Ament2010b}%
  \BibitemOpen
  \bibfield{author}{%
  \bibinfo {author} {\bibfnamefont{L.~J.~P.}\ \bibnamefont{Ament}}, \bibinfo
  {author} {\bibfnamefont{G.}~\bibnamefont{Khaliullin}},\ and\ \bibinfo
  {author} {\bibfnamefont{J.}~\bibnamefont{van~den Brink}},\ }%
  \bibfield{journal}{%
  \bibinfo {journal} {Phys. Rev. B}\ }%
  \textbf{\bibinfo {volume} {84}},\ \bibinfo {pages} {020403} (\bibinfo {year}
  {2011})%
  \bibAnnoteFile{NoStop}{Ament2010b}%
\bibitem{PrivComm}%
  \BibitemOpen
  \bibinfo {note} {G. Ghiringhelli, private communication.}%
  \bibAnnoteFile{Stop}{PrivComm}%
\bibitem{NSLS-II_design_report}%
  \BibitemOpen
  \emph{\bibinfo {title} {Conceptual Design Report National Synchrotron Light
  Source II}}\ (\bibinfo {publisher} {Brookhaven National Laboratory},\
  \bibinfo {address} {Upton (NY), USA},\ \bibinfo {year} {2006})%
  \bibAnnoteFile{NoStop}{NSLS-II_design_report}%
\bibitem{Yu2009}%
  \BibitemOpen
  \bibfield{author}{%
  \bibinfo {author} {\bibfnamefont{G.}~\bibnamefont{Yu}}, \bibinfo {author}
  {\bibfnamefont{Y.}~\bibnamefont{Li}}, \bibinfo {author}
  {\bibfnamefont{E.~M.}\ \bibnamefont{Motoyama}},\ and\ \bibinfo {author}
  {\bibfnamefont{M.}~\bibnamefont{Greven}},\ }%
  \bibfield{journal}{%
  \bibinfo {journal} {Nature Physics}\ }%
  \textbf{\bibinfo {volume} {{5}}},\ \bibinfo {pages} {{873}} (\bibinfo {year}
  {{2009}})%
  \bibAnnoteFile{NoStop}{Yu2009}%
\bibitem{Hoffman2002}%
  \BibitemOpen
  \bibfield{author}{%
  \bibinfo {author} {\bibfnamefont{J.~E.}\ \bibnamefont{Hoffman}}, \bibinfo
  {author} {\bibfnamefont{K.}~\bibnamefont{McElroy}}, \bibinfo {author}
  {\bibfnamefont{D.-H.}\ \bibnamefont{Lee}}, \bibinfo {author}
  {\bibfnamefont{K.~M.}\ \bibnamefont{Lang}}, \bibinfo {author}
  {\bibfnamefont{H.}~\bibnamefont{Eisaki}}, \bibinfo {author}
  {\bibfnamefont{S.}~\bibnamefont{Uchida}},\ and\ \bibinfo {author}
  {\bibfnamefont{J.~C.}\ \bibnamefont{Davis}},\ }%
  \bibfield{journal}{%
  \bibinfo {journal} {Science}\ }%
  \textbf{\bibinfo {volume} {297}},\ \bibinfo {pages} {1148} (\bibinfo {year}
  {2002})%
  \bibAnnoteFile{NoStop}{Hoffman2002}%
\bibitem{McElroy2003}%
  \BibitemOpen
  \bibfield{author}{%
  \bibinfo {author} {\bibfnamefont{K.}~\bibnamefont{McElroy}}, \bibinfo
  {author} {\bibfnamefont{R.}~\bibnamefont{Simmonds}}, \bibinfo {author}
  {\bibfnamefont{J.}~\bibnamefont{Hoffman}}, \bibinfo {author}
  {\bibfnamefont{D.}~\bibnamefont{Lee}}, \bibinfo {author}
  {\bibfnamefont{J.}~\bibnamefont{Orenstein}}, \bibinfo {author}
  {\bibfnamefont{H.}~\bibnamefont{Eisaki}}, \bibinfo {author}
  {\bibfnamefont{S.}~\bibnamefont{Uchida}},\ and\ \bibinfo {author}
  {\bibfnamefont{J.}~\bibnamefont{Davis}},\ }%
  \bibfield{journal}{%
  \bibinfo {journal} {{Nature}}\ }%
  \textbf{\bibinfo {volume} {{422}}},\ \bibinfo {pages} {592} (\bibinfo {year}
  {{2003}})%
  \bibAnnoteFile{NoStop}{McElroy2003}%
\bibitem{Hanaguri2007}%
  \BibitemOpen
  \bibfield{author}{%
  \bibinfo {author} {\bibfnamefont{T.}~\bibnamefont{Hanaguri}}, \bibinfo
  {author} {\bibfnamefont{Y.}~\bibnamefont{Kohsaka}}, \bibinfo {author}
  {\bibfnamefont{J.~C.}\ \bibnamefont{Davis}}, \bibinfo {author}
  {\bibfnamefont{C.}~\bibnamefont{Lupien}}, \bibinfo {author}
  {\bibfnamefont{I.}~\bibnamefont{Yamada}}, \bibinfo {author}
  {\bibfnamefont{M.}~\bibnamefont{Azuma}}, \bibinfo {author}
  {\bibfnamefont{M.}~\bibnamefont{Takano}}, \bibinfo {author}
  {\bibfnamefont{K.}~\bibnamefont{Ohishi}}, \bibinfo {author}
  {\bibfnamefont{M.}~\bibnamefont{Ono}},\ and\ \bibinfo {author}
  {\bibfnamefont{H.}~\bibnamefont{Takagi}},\ }%
  \bibfield{journal}{%
  \bibinfo {journal} {{Nature Physics}}\ }%
  \textbf{\bibinfo {volume} {{3}}},\ \bibinfo {pages} {{865}} (\bibinfo {year}
  {{2007}})%
  \bibAnnoteFile{NoStop}{Hanaguri2007}%
\bibitem{Kohsaka2008}%
  \BibitemOpen
  \bibfield{author}{%
  \bibinfo {author} {\bibfnamefont{Y.}~\bibnamefont{Kohsaka}}, \bibinfo
  {author} {\bibfnamefont{C.}~\bibnamefont{Taylor}}, \bibinfo {author}
  {\bibfnamefont{P.}~\bibnamefont{Wahl}}, \bibinfo {author}
  {\bibfnamefont{A.}~\bibnamefont{Schmidt}}, \bibinfo {author}
  {\bibfnamefont{J.}~\bibnamefont{Lee}}, \bibinfo {author}
  {\bibfnamefont{K.}~\bibnamefont{Fujita}}, \bibinfo {author}
  {\bibfnamefont{J.~W.}\ \bibnamefont{Alldredge}}, \bibinfo {author}
  {\bibfnamefont{K.}~\bibnamefont{McElroy}}, \bibinfo {author}
  {\bibfnamefont{J.}~\bibnamefont{Lee}}, \bibinfo {author}
  {\bibfnamefont{H.}~\bibnamefont{Eisaki}}, \bibinfo {author}
  {\bibfnamefont{S.}~\bibnamefont{Uchida}}, \bibinfo {author}
  {\bibfnamefont{D.~H.}\ \bibnamefont{Lee}},\ and\ \bibinfo {author}
  {\bibfnamefont{J.~C.}\ \bibnamefont{Davis}},\ }%
  \bibfield{journal}{%
  \bibinfo {journal} {{Nature}}\ }%
  \textbf{\bibinfo {volume} {{454}}},\ \bibinfo {pages} {{1072}} (\bibinfo
  {year} {{2008}})%
  \bibAnnoteFile{NoStop}{Kohsaka2008}%
\bibitem{Hanaguri2009}%
  \BibitemOpen
  \bibfield{author}{%
  \bibinfo {author} {\bibfnamefont{T.}~\bibnamefont{Hanaguri}}, \bibinfo
  {author} {\bibfnamefont{Y.}~\bibnamefont{Kohsaka}}, \bibinfo {author}
  {\bibfnamefont{M.}~\bibnamefont{Ono}}, \bibinfo {author}
  {\bibfnamefont{M.}~\bibnamefont{Maltseva}}, \bibinfo {author}
  {\bibfnamefont{P.}~\bibnamefont{Coleman}}, \bibinfo {author}
  {\bibfnamefont{I.}~\bibnamefont{Yamada}}, \bibinfo {author}
  {\bibfnamefont{M.}~\bibnamefont{Azuma}}, \bibinfo {author}
  {\bibfnamefont{M.}~\bibnamefont{Takano}}, \bibinfo {author}
  {\bibfnamefont{K.}~\bibnamefont{Ohishi}},\ and\ \bibinfo {author}
  {\bibfnamefont{H.}~\bibnamefont{Takagi}},\ }%
  \bibfield{journal}{%
  \bibinfo {journal} {{Science}}\ }%
  \textbf{\bibinfo {volume} {{323}}},\ \bibinfo {pages} {{923}} (\bibinfo
  {year} {{2009}})%
  \bibAnnoteFile{NoStop}{Hanaguri2009}%
\bibitem{Hanke12}%
  \BibitemOpen
  \bibfield{author}{%
  \bibinfo {author} {\bibfnamefont{T.}~\bibnamefont{H\"anke}}, \bibinfo
  {author} {\bibfnamefont{S.}~\bibnamefont{Sykora}}, \bibinfo {author}
  {\bibfnamefont{R.}~\bibnamefont{Schlegel}}, \bibinfo {author}
  {\bibfnamefont{D.}~\bibnamefont{Baumann}}, \bibinfo {author}
  {\bibfnamefont{L.}~\bibnamefont{Harnagea}}, \bibinfo {author}
  {\bibfnamefont{S.}~\bibnamefont{Wurmehl}}, \bibinfo {author}
  {\bibfnamefont{M.}~\bibnamefont{Daghofer}}, \bibinfo {author}
  {\bibfnamefont{B.}~\bibnamefont{B\"uchner}}, \bibinfo {author}
  {\bibfnamefont{J.}~\bibnamefont{van~den Brink}},\ and\ \bibinfo {author}
  {\bibfnamefont{C.}~\bibnamefont{Hess}},\ }%
  \bibfield{journal}{%
  \bibinfo {journal} {Phys. Rev. Lett.}\ }%
  \textbf{\bibinfo {volume} {108}},\ \bibinfo {pages} {127001} (\bibinfo {year}
  {2012})%
  \bibAnnoteFile{NoStop}{Hanke12}%
\bibitem{Sykora11}%
  \BibitemOpen
  \bibfield{author}{%
  \bibinfo {author} {\bibfnamefont{S.}~\bibnamefont{Sykora}}\ and\ \bibinfo
  {author} {\bibfnamefont{P.}~\bibnamefont{Coleman}},\ }%
  \bibfield{journal}{%
  \bibinfo {journal} {Phys. Rev. B}\ }%
  \textbf{\bibinfo {volume} {84}},\ \bibinfo {pages} {054501} (\bibinfo {year}
  {2011})%
  \bibAnnoteFile{NoStop}{Sykora11}%
\bibitem{Haverkort2010}%
  \BibitemOpen
  \bibfield{author}{%
  \bibinfo {author} {\bibfnamefont{M.~W.}\ \bibnamefont{Haverkort}},\ }%
  \bibfield{journal}{%
  \bibinfo {journal} {Phys. Rev. Lett.}\ }%
  \textbf{\bibinfo {volume} {105}},\ \bibinfo {pages} {167404} (\bibinfo {year}
  {2010})%
  \bibAnnoteFile{NoStop}{Haverkort2010}%
\bibitem{Marra2012}%
  \BibitemOpen
  \bibfield{author}{%
  \bibinfo {author} {\bibfnamefont{P.}~\bibnamefont{Marra}}, \bibinfo {author}
  {\bibfnamefont{K.}~\bibnamefont{Wohlfeld}},\ and\ \bibinfo {author}
  {\bibfnamefont{J.}~\bibnamefont{van~den Brink}},\ }%
  \bibfield{journal}{%
  \bibinfo {journal} {Phys. Rev. Lett.}\ }%
  \textbf{\bibinfo {volume} {109}},\ \bibinfo {pages} {117401} (\bibinfo {year}
  {2012})%
  \bibAnnoteFile{NoStop}{Marra2012}%
\bibitem{Note1}%
  \BibitemOpen
  \bibinfo {note} {Note that the quantization axis for the spin operator in
  RIXS is not arbitrary and, e.g., in cuprates it is perpendicular to the plane
  in which the 3$d_{x^2-y^2}$ orbital is located~\cite{Ament2009a,
  Haverkort2010, Marra2012}.}%
  \bibAnnoteFile{Stop}{Note1}%
\bibitem{Braicovich2010}%
  \BibitemOpen
  \bibfield{author}{%
  \bibinfo {author} {\bibfnamefont{L.}~\bibnamefont{Braicovich}}, \bibinfo
  {author} {\bibfnamefont{M.}~\bibnamefont{Moretti~Sala}}, \bibinfo {author}
  {\bibfnamefont{L.~J.~P.}\ \bibnamefont{Ament}}, \bibinfo {author}
  {\bibfnamefont{V.}~\bibnamefont{Bisogni}}, \bibinfo {author}
  {\bibfnamefont{M.}~\bibnamefont{Minola}}, \bibinfo {author}
  {\bibfnamefont{G.}~\bibnamefont{Balestrino}}, \bibinfo {author}
  {\bibfnamefont{D.}~\bibnamefont{Di~Castro}}, \bibinfo {author}
  {\bibfnamefont{G.~M.}\ \bibnamefont{De~Luca}}, \bibinfo {author}
  {\bibfnamefont{M.}~\bibnamefont{Salluzzo}}, \bibinfo {author}
  {\bibfnamefont{G.}~\bibnamefont{Ghiringhelli}},\ and\ \bibinfo {author}
  {\bibfnamefont{J.}~\bibnamefont{van~den Brink}},\ }%
  \bibfield{journal}{%
  \bibinfo {journal} {Phys. Rev. B}\ }%
  \textbf{\bibinfo {volume} {81}},\ \bibinfo {pages} {174533} (\bibinfo {year}
  {2010})%
  \bibAnnoteFile{NoStop}{Braicovich2010}%
\bibitem{Kaneshita2011}%
  \BibitemOpen
  \bibfield{author}{%
  \bibinfo {author} {\bibfnamefont{E.}~\bibnamefont{Kaneshita}}, \bibinfo
  {author} {\bibfnamefont{K.}~\bibnamefont{Tsutsui}},\ and\ \bibinfo {author}
  {\bibfnamefont{T.}~\bibnamefont{Tohyama}},\ }%
  \bibfield{journal}{%
  \bibinfo {journal} {Phys. Rev. B}\ }%
  \textbf{\bibinfo {volume} {84}},\ \bibinfo {pages} {020511} (\bibinfo {year}
  {2011})%
  \bibAnnoteFile{NoStop}{Kaneshita2011}%
\bibitem{Kim2012a}%
  \BibitemOpen
  \bibfield{author}{%
  \bibinfo {author} {\bibfnamefont{J.}~\bibnamefont{Kim}}, \bibinfo {author}
  {\bibfnamefont{D.}~\bibnamefont{Casa}}, \bibinfo {author}
  {\bibfnamefont{M.~H.}\ \bibnamefont{Upton}}, \bibinfo {author}
  {\bibfnamefont{T.}~\bibnamefont{Gog}}, \bibinfo {author}
  {\bibfnamefont{Y.-J.}\ \bibnamefont{Kim}}, \bibinfo {author}
  {\bibfnamefont{J.~F.}\ \bibnamefont{Mitchell}}, \bibinfo {author}
  {\bibfnamefont{M.}~\bibnamefont{van Veenendaal}}, \bibinfo {author}
  {\bibfnamefont{M.}~\bibnamefont{Daghofer}}, \bibinfo {author}
  {\bibfnamefont{J.}~\bibnamefont{van~den Brink}}, \bibinfo {author}
  {\bibfnamefont{G.}~\bibnamefont{Khaliullin}},\ and\ \bibinfo {author}
  {\bibfnamefont{B.~J.}\ \bibnamefont{Kim}},\ }%
  \bibfield{journal}{%
  \bibinfo {journal} {Phys. Rev. Lett.}\ }%
  \textbf{\bibinfo {volume} {108}},\ \bibinfo {pages} {177003} (\bibinfo {year}
  {2012})%
  \bibAnnoteFile{NoStop}{Kim2012a}%
\bibitem{Andersen2005}%
  \BibitemOpen
  \bibfield{author}{%
  \bibinfo {author} {\bibfnamefont{B.~M.}\ \bibnamefont{Andersen}}\ and\
  \bibinfo {author} {\bibfnamefont{P.}~\bibnamefont{Hedeg\aa{}rd}},\ }%
  \bibfield{journal}{%
  \bibinfo {journal} {Phys. Rev. Lett.}\ }%
  \textbf{\bibinfo {volume} {95}},\ \bibinfo {pages} {037002} (\bibinfo {year}
  {2005})%
  \bibAnnoteFile{NoStop}{Andersen2005}%
\bibitem{Fischer2007}%
  \BibitemOpen
  \bibfield{author}{%
  \bibinfo {author} {\bibfnamefont{O.}~\bibnamefont{Fischer}}, \bibinfo
  {author} {\bibfnamefont{M.}~\bibnamefont{Kugler}}, \bibinfo {author}
  {\bibfnamefont{I.}~\bibnamefont{Maggio-Aprile}}, \bibinfo {author}
  {\bibfnamefont{C.}~\bibnamefont{Berthod}},\ and\ \bibinfo {author}
  {\bibfnamefont{C.}~\bibnamefont{Renner}},\ }%
  \bibfield{journal}{%
  \bibinfo {journal} {Rev. Mod. Phys.}\ }%
  \textbf{\bibinfo {volume} {79}},\ \bibinfo {pages} {353} (\bibinfo {year}
  {2007})%
  \bibAnnoteFile{NoStop}{Fischer2007}%
\bibitem{Kee1998}%
  \BibitemOpen
  \bibfield{author}{%
  \bibinfo {author} {\bibfnamefont{H.-Y.}\ \bibnamefont{Kee}}\ and\ \bibinfo
  {author} {\bibfnamefont{C.~M.}\ \bibnamefont{Varma}},\ }%
  \bibfield{journal}{%
  \bibinfo {journal} {Phys. Rev. B}\ }%
  \textbf{\bibinfo {volume} {58}},\ \bibinfo {pages} {15035} (\bibinfo {year}
  {1998})%
  \bibAnnoteFile{NoStop}{Kee1998}%
\bibitem{Kee1999}%
  \BibitemOpen
  \bibfield{author}{%
  \bibinfo {author} {\bibfnamefont{H.-Y.}\ \bibnamefont{Kee}}\ and\ \bibinfo
  {author} {\bibfnamefont{Y.~B.}\ \bibnamefont{Kim}},\ }%
  \bibfield{journal}{%
  \bibinfo {journal} {Phys. Rev. B}\ }%
  \textbf{\bibinfo {volume} {59}},\ \bibinfo {pages} {4470} (\bibinfo {year}
  {1999})%
  \bibAnnoteFile{NoStop}{Kee1999}%
\bibitem{Voo2000}%
  \BibitemOpen
  \bibfield{author}{%
  \bibinfo {author} {\bibfnamefont{K.-K.}\ \bibnamefont{Voo}}, \bibinfo
  {author} {\bibfnamefont{W.~C.}\ \bibnamefont{Wu}}, \bibinfo {author}
  {\bibfnamefont{J.-X.}\ \bibnamefont{Li}},\ and\ \bibinfo {author}
  {\bibfnamefont{T.~K.}\ \bibnamefont{Lee}},\ }%
  \bibfield{journal}{%
  \bibinfo {journal} {Phys. Rev. B}\ }%
  \textbf{\bibinfo {volume} {61}},\ \bibinfo {pages} {9095} (\bibinfo {year}
  {2000})%
  \bibAnnoteFile{NoStop}{Voo2000}%
\bibitem{Kao2005}%
  \BibitemOpen
  \bibfield{author}{%
  \bibinfo {author} {\bibfnamefont{Y.-J.}\ \bibnamefont{Kao}}\ and\ \bibinfo
  {author} {\bibfnamefont{H.-Y.}\ \bibnamefont{Kee}},\ }%
  \bibfield{journal}{%
  \bibinfo {journal} {Phys. Rev. B}\ }%
  \textbf{\bibinfo {volume} {72}},\ \bibinfo {pages} {024502} (\bibinfo {year}
  {2005})%
  \bibAnnoteFile{NoStop}{Kao2005}%
\bibitem{Schrieffer1989}%
  \BibitemOpen
  \bibfield{author}{%
  \bibinfo {author} {\bibfnamefont{J.~R.}\ \bibnamefont{Schrieffer}}, \bibinfo
  {author} {\bibfnamefont{X.~G.}\ \bibnamefont{Wen}},\ and\ \bibinfo {author}
  {\bibfnamefont{S.~C.}\ \bibnamefont{Zhang}},\ }%
  \bibfield{journal}{%
  \bibinfo {journal} {Phys. Rev. B}\ }%
  \textbf{\bibinfo {volume} {39}},\ \bibinfo {pages} {11663} (\bibinfo {year}
  {1989})%
  \bibAnnoteFile{NoStop}{Schrieffer1989}%
\bibitem{Kuroki2001}%
  \BibitemOpen
  \bibfield{author}{%
  \bibinfo {author} {\bibfnamefont{K.}~\bibnamefont{Kuroki}}\ and\ \bibinfo
  {author} {\bibfnamefont{R.}~\bibnamefont{Arita}},\ }%
  \bibfield{journal}{%
  \bibinfo {journal} {Phys. Rev. B}\ }%
  \textbf{\bibinfo {volume} {64}},\ \bibinfo {pages} {024501} (\bibinfo {year}
  {2001})%
  \bibAnnoteFile{NoStop}{Kuroki2001}%
\bibitem{Trugman1990}%
  \BibitemOpen
  \bibfield{author}{%
  \bibinfo {author} {\bibfnamefont{S.~A.}\ \bibnamefont{Trugman}},\ }%
  \bibfield{journal}{%
  \bibinfo {journal} {Phys. Rev. B}\ }%
  \textbf{\bibinfo {volume} {42}},\ \bibinfo {pages} {6612} (\bibinfo {year}
  {1990})%
  \bibAnnoteFile{NoStop}{Trugman1990}%
\bibitem{Becca2000}%
  \BibitemOpen
  \bibfield{author}{%
  \bibinfo {author} {\bibfnamefont{F.}~\bibnamefont{Becca}}, \bibinfo {author}
  {\bibfnamefont{L.}~\bibnamefont{Capriotti}}, \bibinfo {author}
  {\bibfnamefont{S.}~\bibnamefont{Sorella}},\ and\ \bibinfo {author}
  {\bibfnamefont{A.}~\bibnamefont{Parola}},\ }%
  \bibfield{journal}{%
  \bibinfo {journal} {Phys. Rev. B}\ }%
  \textbf{\bibinfo {volume} {62}},\ \bibinfo {pages} {15277} (\bibinfo {year}
  {2000})%
  \bibAnnoteFile{NoStop}{Becca2000}%
\bibitem{Chao1978}%
  \BibitemOpen
  \bibfield{author}{%
  \bibinfo {author} {\bibfnamefont{K.~A.}\ \bibnamefont{Chao}}, \bibinfo
  {author} {\bibfnamefont{J.}~\bibnamefont{Spa\l{}ek}},\ and\ \bibinfo {author}
  {\bibfnamefont{A.~M.}\ \bibnamefont{Ole\ifmmode~\acute{s}\else \'{s}\fi{}}},\
  }%
  \bibfield{journal}{%
  \bibinfo {journal} {Phys. Rev. B}\ }%
  \textbf{\bibinfo {volume} {18}},\ \bibinfo {pages} {3453} (\bibinfo {year}
  {1978})%
  \bibAnnoteFile{NoStop}{Chao1978}%
\bibitem{Putikka1992}%
  \BibitemOpen
  \bibfield{author}{%
  \bibinfo {author} {\bibfnamefont{W.~O.}\ \bibnamefont{Putikka}}, \bibinfo
  {author} {\bibfnamefont{M.~U.}\ \bibnamefont{Luchini}},\ and\ \bibinfo
  {author} {\bibfnamefont{T.~M.}\ \bibnamefont{Rice}},\ }%
  \bibfield{journal}{%
  \bibinfo {journal} {Phys. Rev. Lett.}\ }%
  \textbf{\bibinfo {volume} {68}},\ \bibinfo {pages} {538} (\bibinfo {year}
  {1992})%
  \bibAnnoteFile{NoStop}{Putikka1992}%
\bibitem{Tandon1999}%
  \BibitemOpen
  \bibfield{author}{%
  \bibinfo {author} {\bibfnamefont{A.}~\bibnamefont{Tandon}}, \bibinfo {author}
  {\bibfnamefont{Z.}~\bibnamefont{Wang}},\ and\ \bibinfo {author}
  {\bibfnamefont{G.}~\bibnamefont{Kotliar}},\ }%
  \bibfield{journal}{%
  \bibinfo {journal} {Phys. Rev. Lett.}\ }%
  \textbf{\bibinfo {volume} {83}},\ \bibinfo {pages} {2046} (\bibinfo {year}
  {1999})%
  \bibAnnoteFile{NoStop}{Tandon1999}%
\bibitem{Cosentini1998}%
  \BibitemOpen
  \bibfield{author}{%
  \bibinfo {author} {\bibfnamefont{A.~C.}\ \bibnamefont{Cosentini}}, \bibinfo
  {author} {\bibfnamefont{M.}~\bibnamefont{Capone}}, \bibinfo {author}
  {\bibfnamefont{L.}~\bibnamefont{Guidoni}},\ and\ \bibinfo {author}
  {\bibfnamefont{G.~B.}\ \bibnamefont{Bachelet}},\ }%
  \bibfield{journal}{%
  \bibinfo {journal} {Phys. Rev. B}\ }%
  \textbf{\bibinfo {volume} {58}},\ \bibinfo {pages} {R14685} (\bibinfo {year}
  {1998})%
  \bibAnnoteFile{NoStop}{Cosentini1998}%
\bibitem{Sykora2009}%
  \BibitemOpen
  \bibfield{author}{%
  \bibinfo {author} {\bibfnamefont{S.}~\bibnamefont{Sykora}}\ and\ \bibinfo
  {author} {\bibfnamefont{K.~W.}\ \bibnamefont{Becker}},\ }%
  \bibfield{journal}{%
  \bibinfo {journal} {Phys. Rev. B}\ }%
  \textbf{\bibinfo {volume} {80}},\ \bibinfo {pages} {014511} (\bibinfo {year}
  {2009})%
  \bibAnnoteFile{NoStop}{Sykora2009}%
\bibitem{Fazekas}%
  \BibitemOpen
  \bibfield{author}{%
  \bibinfo {author} {\bibfnamefont{P.}~\bibnamefont{Fazekas}},\ }%
  \emph{\bibinfo {title} {{Lecture Notes on Electron Correlation and Magnetism
  (Series in Modern Condensed Matter Physics, Vol. 5)}}}\ (\bibinfo {year}
  {1999})%
  \bibAnnoteFile{NoStop}{Fazekas}%
\end{thebibliography}
\end{document}